\documentclass[twocolumn,preprintnumbers,superscriptaddress,nofootinbib,aps,prd,floatfix]{revtex4}
\pdfoutput=1
\usepackage[utf8]{inputenc}
\usepackage{footmisc,multirow}
\usepackage{xspace}
\usepackage{amsmath,amssymb,slashed,bbm}
\usepackage{braket}
\usepackage{booktabs}
\usepackage{graphicx}
\usepackage{url}
\usepackage[breaklinks=true]{hyperref}
\hypersetup{
     colorlinks   	= true,
     citecolor    	= LightSeaGreen,
     linkcolor 	= black,
     menucolor	= black,
}
\usepackage{hhline,multirow}
\usepackage{adjustbox}
\usepackage{enumitem}
\allowdisplaybreaks
\usepackage{subfigure}
\usepackage{array}

\numberwithin{equation}{section}
\numberwithin{table}{section}

\makeatletter
\@addtoreset{footnote}{section}
\makeatother

\usepackage[dvipsnames,svgnames,x11names,hyperref]{xcolor}
\definecolor{comment}{rgb}{0,0.3,0}
\definecolor{identifier}{rgb}{0.0,0,0.3}
\usepackage{listings}
\newcommand{\cp}{\ensuremath{\mathcal{CP}}\xspace}

\newcommand{\gev}{\ensuremath{\text{G}\mspace{0.2mu}\text{e}\mspace{-1mu}\text{V}}\xspace}
\newcommand{\tev}{\ensuremath{\text{T}\mspace{0.2mu}\text{e}\mspace{-1mu}\text{V}}\xspace}

\newcommand{\lag}[1]{\ensuremath{\mathcal{L}_\mathrm{#1}}}
\newcommand{\ord}[1]{\ensuremath{\mathcal{O}(\mathrm{#1})}}
\newcommand{\summ}[1]{\ensuremath{\sum_{\substack{#1}} }}
\newcommand{\op}[2][]{\ensuremath{\mathcal{O}_{#2}^{#1}}}
\newcommand{\co}[2][]{\ensuremath{c_{#2}^{#1}}}
\newcommand{\cb}[2][]{\ensuremath{\bar{c}_{#2}^{#1}}}

\newcommand{\D}[1]{\ensuremath{D=#1}}

\newcommand{\fb}{\ensuremath{\text{fb}}\xspace}
\newcommand{\ifb}{\ensuremath{\text{fb}^{-1}}\xspace}
\newcommand{\iab}{\ensuremath{\text{ab}^{-1}}\xspace}
\newcommand{\pb}{\ensuremath{\text{pb}}\xspace}
\newcommand{\ttbar}{\ensuremath{t\bar{t}}\xspace}

\newcommand{\ttz}{\ensuremath{t\bar{t}Z}\xspace}
\newcommand{\ep}{\ensuremath{e^+e^-}\xspace}

\newcommand{\sqrts}{\ensuremath{\sqrt{s}}\xspace}

\graphicspath{{figs/}}

\begin{document}

\title{Top quark electroweak couplings at future lepton colliders}

\begin{abstract}
We perform a comparative study of the reach of future \ep collider options for the scale of non-resonant new physics effects in the top quark sector, phrased in the language of higher-dimensional operators. Our focus is on the electroweak top quark pair production process $\ep \to Z^*/\gamma \to \ttbar $, and we study benchmark scenarios at the ILC and CLIC. We find that both are able to constrain mass scales up to the few \tev range in the most sensitive cases, improving by orders of magnitude on the forecasted capabilities of the LHC. We discuss the role played by observables such as forward-backward asymmetries, and making use of different beam polarisation settings, and highlight the possibility of lifting a degeneracy in the allowed parameter space by combining top observables with precision $Z$-pole measurements from LEP1.
\end{abstract}

\author{Christoph Englert} \email{christoph.englert@glasgow.ac.uk}
\affiliation{School of Physics and Astronomy, University of
  Glasgow,\\Glasgow, G12 8QQ, United Kingdom\\[0.2cm]}

\author{Michael Russell} \email{m.russell.2@research.gla.ac.uk}   
\affiliation{School of Physics and Astronomy, University of
  Glasgow,\\Glasgow, G12 8QQ, United Kingdom\\[0.2cm]}

\maketitle

\section{Introduction}
\label{sec:intro}
Studies of the couplings of the top quark form one of the cornerstones of the analysis program pursued at the LHC, where top quark related processes, most notably \ttbar production, are ubiquitous. Whilst improving the precision of Standard Model predictions is important in its own right, a key motivation for the detailed study of the top quark is to understand its potential role in the physics behind electroweak symmetry breaking, which has yet to be elucidated. As the only fermion with a Yukawa coupling $y_t \simeq 1 $, the top is a pertinent place to look for the signatures of new physics at the \tev scale. Motivated scenarios range from compositeness theories~(e.g.~\cite{Georgi:1994ha,Contino:2003ve,Agashe:2004rs}) all the way to supersymmetry (see \cite{Nilles:1983ge} for a review).

If the new physics that couples to the top quark is heavy and/or weakly coupled, an appropriate formulation of the effects of new physics is to view the SM as an effective theory, where the non-standard couplings are parameterised by operators of dimension $D>4$. The leading effects for collider observables typically enter at \D6, 
\begin{equation}
\lag{eff} = \lag{SM}+\summ{i}\frac{\co[(6)]{i}\op[(6)]{i}}{\Lambda^2}+\ldots
\label{eqn:eftexpn}
\end{equation}
It is also typically justified to truncate the EFT expansion\footnote{For counterexamples to this statement, see for example Ref.~\cite{Contino:2016jqw} and references therein.} at this order. Global fits of top quark couplings to data from ATLAS and CMS are already well established (see e.g. Refs.~\cite{Buckley:2015nca,Buckley:2015lku,Bylund:2016phk,Durieux:2014xla,Rontsch:2014cca,Rosello:2015sck,Schulze:2016qas} for a representative sample as well as \cite{deBlas:2015aea} for a discussion of electroweak precision effects), and a similar effort is well underway in Higgs physics, where the associated limits on Wilson coefficients are typically statistics limited.

\begin{figure*}[t!]
  \begin{minipage}[c]{0.6\textwidth}
  \begin{center}
    \includegraphics[width=\textwidth]{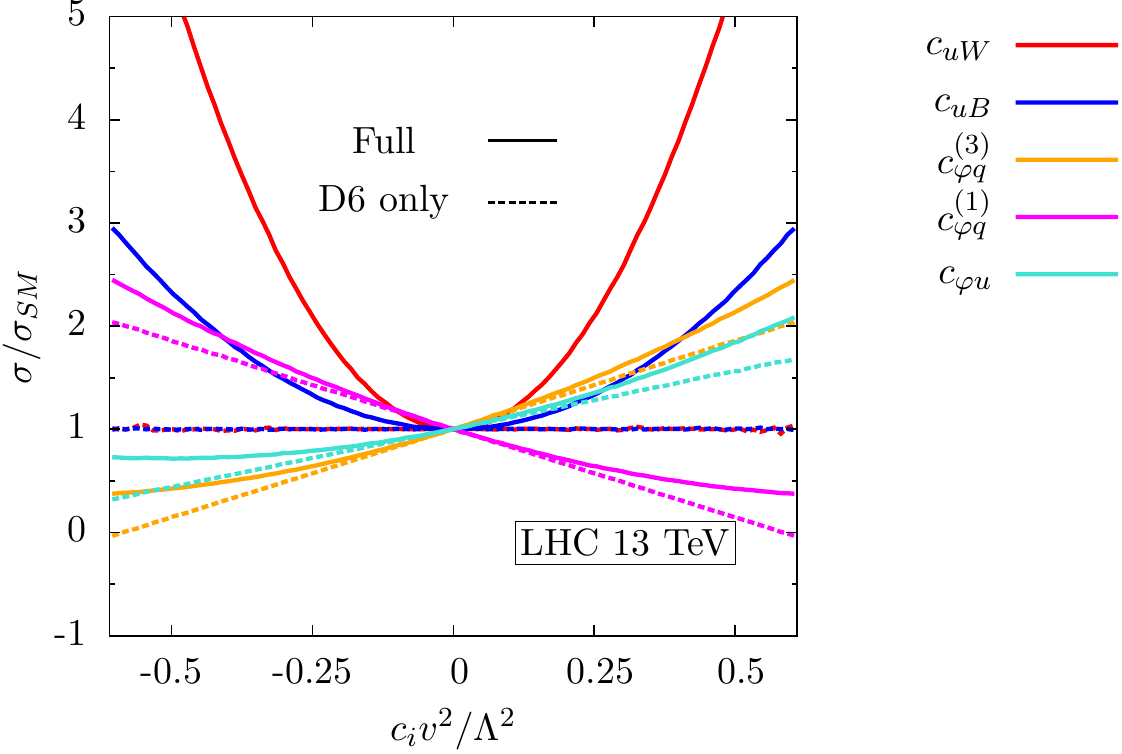}
    \end{center}
  \end{minipage} 
\hspace{-60pt}
  \begin{minipage}[r]{0.45\textwidth}
  \vspace{3.cm}
    \caption{Ratio of the full SM $pp \to \ttz$ cross-section with the operators of Eq.~\eqref{eqn:ttzops} switched on individually to the NLO Standard Model estimate. The dashed lines show the contribution from the interference term, and the solid lines show the full dependence.\label{fig:ttzxsec}}
  \end{minipage}
\end{figure*}

%
Still, despite the impressive statistical sample of top quark data that enters these fits, the subsequent direct bounds on the operator coefficients, and, by extension, the scale of new physics that would generate those operators, are rather weak~\cite{Buckley:2015lku}. There are few top quark measurements at the LHC that can be considered ``precision observables" (helicity fractions in top decays are an exception~\cite{AguilarSaavedra:2006fy}). By dimensional analysis, the strength of the interference of these operators with energy \sqrts typically scales as $s/\Lambda^2$, and the areas of phase space that are most sensitive are plagued by correlated experimental and theoretical systematics. Moreover, the associated weak limits translate into values of $\Lambda$ that are probed by the high energy bins of the measurement, bringing into question the validity of the truncated EFT description~\cite{Contino:2016jqw} and care needs to be taken when combining measurements of different exclusive energy ranges of a binned distribution~\cite{Englert:2014cva}. Inclusive cross-sections, being typically dominated by the threshold region $\sqrts \sim (2)m_t $, are under more theoretical control, but bring far less sensitivity.

Lepton colliders are not vulnerable to either of these problems. Firstly, there is excellent control over the hard scale of the interaction $\sqrt{s}$, so one can always ensure that the limits on the \D6 operators are consistent with a well-behaved EFT expansion.\footnote{Consistently improving the perturbative precision within the dimension 6 framework, however, makes the truncation of the perturbative series necessary as corrections to $(D=6)^2$ operators will typically require unaccounted $D=8$ counterterms.} Secondly, the theoretical uncertainties from Standard Model calculations are much smaller: there are no PDFs, and the current state of the art precision for \ttbar production is N$^3$LO QCD at fixed-order~\cite{Kiyo:2009gb}, and NNLO+NNLL including threshold resummation, which bring SM scale uncertainty variation bands to the percent level~\cite{Chen:2016zbz}. 

The physics case for a \ep collider is by now well-established. The principal motivation is to perform a detailed precision study of the couplings of the Higgs boson in the much cleaner environment that a lepton collider affords, which will bring Higgs coupling measurements to an accuracy that will not be challenged by the LHC, even after it collects 3~\iab of data~\cite{Klute:2013cx,Khanpour:2017cfq}. 

The electroweak couplings of the top quark are also clearly within the remit of such a collider. Currently, the only handle on top quark electroweak couplings from the LHC is through the associated production $pp \to \ttbar V$ where $V \in \{Z,W,\gamma\}$. Whilst measurements of these processes are now approaching the 5$\sigma$ level, the pull that they have on a global fit is small~\cite{Buckley:2015lku}. Measurements of electroweak single top production bring stronger bounds, but are sensitive to a smaller subset of operators. 

At a lepton collider, on the other hand, the process $\ep \to Z^*/\gamma \to \ttbar$ is extremely sensitive to top electroweak couplings. While the overall rate is more modest than at the LHC due to the parametric $\alpha_{EW}/\alpha_s$ and $s$-channel suppression, the process is essentially background-free, and would constitute the first true precision probe of the electroweak sector of the top quark, and open up a new avenue for top quark couplings, complementary to the well-studied top QCD interactions.

Several studies of the prospects for improvement of top measurements at future colliders have already been undertaken (see for example Refs.~\cite{Vos:2017ses,Vos:2016til,Amjad:2015mma,Rontsch:2015una,Coleppa:2017rgb,Cao:2015qta}), in particular for the proposed International Linear Collider (ILC), but none have explicitly quantified the gain in the constraints on the top electroweak sector of the SMEFT, nor provided a comparative study of different collider options. The purpose of this paper is to provide such a study.

In Sec.~\ref{sec:topew} we discuss the $\ep \to Z^*/\gamma \to \ttbar$ process in the Standard Model EFT, and the \D6 operators that generate interference with the Standard Model. To motivate the study, in Sec.~\ref{sec:had} we discuss the present status of constraints on top electroweak couplings, and discuss projections for the lifetime of the LHC. The rest of the paper is devoted to lepton collider projections. In Sec.~\ref{sec:ilc} we discuss ILC constraints, based on the H-20 running scenario. In Sec.~\ref{sec:clic}, we focus on prospects from CLIC, comparing the potential of the two future collider projections. In Sec.~\ref{sec:lep} we analyse the complementarity between the bounds derived from \ttbar measurements at a future collider and $Z$-pole measurements, before summarising in Sec.~\ref{sec:conc}.

\section{Top electroweak couplings}
\label{sec:topew}
In the Standard Model, the electroweak \ttz coupling is given by the vector-axial-vector coupling 
\begin{equation}
\lag{ttZ} = e \bar t\left[ \gamma^\mu(v_t-\gamma_5 a_t)\right] t Z_\mu
\label{eqn:ttz}
\end{equation}
where
\begin{equation}
\begin{split}
v_t &= \frac{T_t^3 - 2Q_t \sin^2 \theta_W}{2\sin\theta_W\cos\theta_W} \simeq 0.24, \\
a_t &= \frac{T_t^3}{2\sin\theta_W\cos\theta_W} \simeq 0.60.
\end{split}
\end{equation}
To capture effects beyond the SM in this Lagrangian there are two approaches: one can write down anomalous couplings for the \ttz vertex, such that \lag{ttZ} receives a term
\begin{equation}
\begin{split}
\Delta \lag{ttZ} =  e \bar t & \left[  \vphantom{\frac{i\sigma^{\mu\nu}q_\nu}{2M_Z}} (\gamma^\mu(C_{1V} + \gamma_5C_{1A}) \right. \\
			  +  & \left. \frac{i\sigma^{\mu\nu}q_\nu}{2M_Z}(C_{2V} +\gamma_5C_{2A}) \right]t Z_\mu ,
\end{split}
\label{eqn:anomcoups}
\end{equation}
where $q = p_t -p_{\bar t}$. While this has the advantage of elucidating the various spin structures that can impact the \ttz vertex, it has the drawback that it does not allow for a simple power counting of which anomalous couplings would have the strongest effect. For example, the coefficient $C_{2A}$ is zero in the Standard Model, so that any corrections to it come solely from new physics contributions, which should be smaller than couplings that have SM interference. 

To augment this description, one can instead supplement Eq.~\eqref{eqn:ttz} with higher-dimensional operators. At leading order in the SMEFT, the list of operators that generate modifications to the \ttz vertex is, expressed in the basis and notation of Ref.~\cite{Grzadkowski:2010es}:
\begin{equation}
\begin{split}
\op{uW} &= (\bar Q\sigma^{\mu\nu}u)\tau^I\tilde \varphi W^I_{\mu\nu} \\
\op{uB}  &= (\bar Q\sigma^{\mu\nu}u)\tilde \varphi B_{\mu\nu} \\
\op[(3)]{\varphi q} &= (\varphi^\dagger i\overleftrightarrow{D^I_\mu} \varphi)(\bar Q \tau^I \gamma^\mu Q) \\
\op[(1)]{\varphi q} &= (\varphi^\dagger i\overleftrightarrow{D_\mu} \varphi)(\bar Q\gamma^\mu Q) \\
\op{\varphi u} &= (\varphi^\dagger i\overleftrightarrow{D_\mu} \varphi)(\bar u \gamma^\mu u) . \\
\end{split}
\label{eqn:ttzops}
\end{equation}
The dictionary between the \D6 operators of Eq.~\eqref{eqn:ttzops} and the anomalous couplings of Eq.~\eqref{eqn:anomcoups} is
\begin{equation}
\begin{split}
C_{1V} &= \frac{v^2}{\Lambda^2}\Re\left[\co[(3)]{\varphi q}-\co[(1)]{\varphi q}-\co{\varphi u}\right]^{33} \\
C_{1A} &= \frac{v^2}{\Lambda^2}\Re\left[\co[(3)]{\varphi q}-\co[(1)]{\varphi q}+\co{\varphi u}\right]^{33} \\
C_{2V} &= \sqrt{2}\frac{v^2}{\Lambda^2}\Re\left[\cos\theta_W\co{uW}-\sin\theta_W\co{uB}\right]^{33} \\
C_{2A} &= \sqrt{2}\frac{v^2}{\Lambda^2}\Im\left[\cos\theta_W\co{uW}+\sin\theta_W\co{uB}\right]^{33},
\end{split}
\end{equation}
where the superscript 33 denotes that we are considering the 3rd generation only in the fermion bilinears of Eq.~\eqref{eqn:ttzops}. Since \co[(3)]{\varphi q} and \co[(1)]{\varphi q} only appear with an overall opposite sign, we can only constrain the operator $\op[(3)]{\varphi q} - \op[(1)]{\varphi q} \equiv \op {\varphi q}$ from \ttz couplings. We will discuss a method for bounding the two operators independently later in the paper.

\begin{figure}[t!]
\begin{center}
 \includegraphics[width=0.45\textwidth]{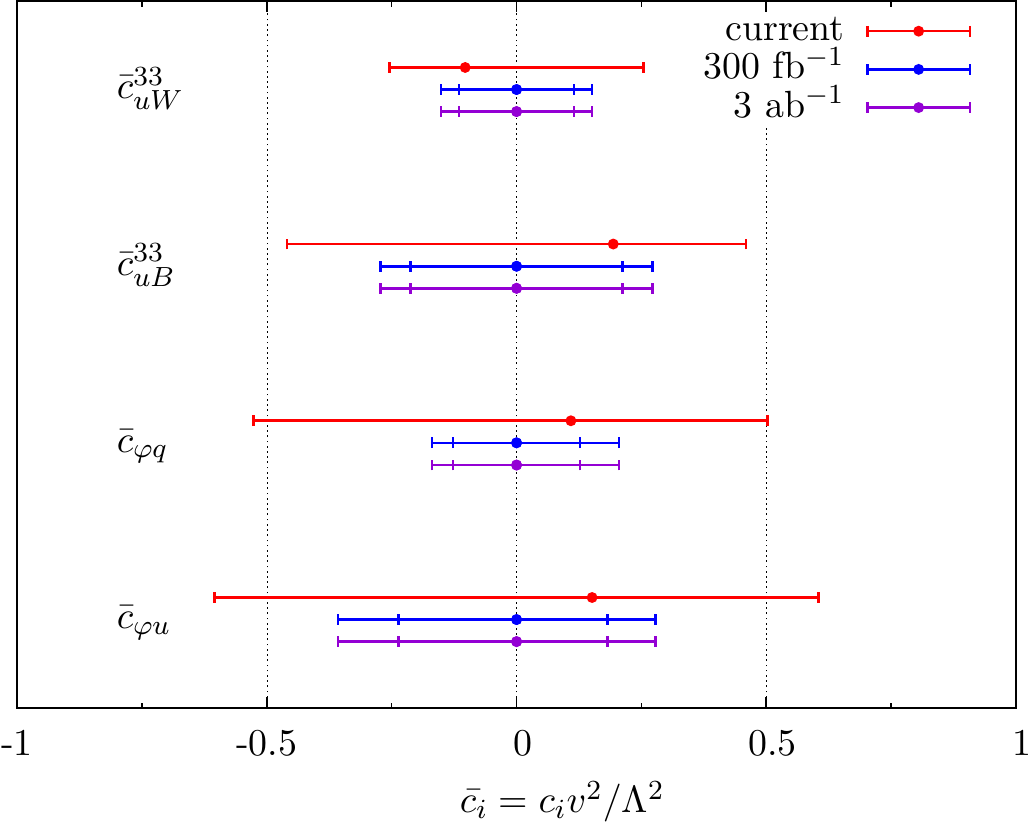}
 \caption{Individual 95\% confidence intervals on the coefficients of the operators of Eq.~\eqref{eqn:ttzops} using the current 13~\tev measurements (red bars). Also shown are the projected constraints using 300~\ifb (blue) and 3~\iab (purple) of SM pseudodata. For the latter two cases, the inner bars show the improvement when theory uncertainties are reduced to 1\%.} \label{fig:ttzconstraints}
 \end{center}
 \end{figure}

$C_{2A}$ is generated by a \cp-odd combination of operators, therefore it does not interfere with SM amplitudes and so its effects are expected to be smaller. Since in this study we are more interested in the absolute mass scales of these operators, we set all Wilson coefficients to be real\footnote{This is a well-motivated assumption considering that \cp-odd components of the operators we consider are very tightly constrained by EDM experiments\cite{Cirigliano:2016njn}.}, however we note that \cp-sensitive observables such as angular distributions can also distinguish the \cp character of the Wilson coefficients. We also assume that the new physics solely impacts the \ttz vertex, so we do not consider operators which modify the $Ze\bar e$ vertex, nor four-fermion operators which can contribute to the $q\bar q \to \ttbar$ or $\ep \to \ttbar$ processes, though we note that including these would in general weaken our bounds (see e.g. Refs.~\cite{Buckley:2015lku,Rosello:2015sck} for constraints on the former and Ref.~\cite{deBlas:2015aea} for the latter).

\section{Analysis setup}
\label{sec:analysis}
We implemented  the operators of Eq.~\eqref{eqn:ttzops} in a {\sc{FeynRules}}~\cite{Alloul:2013bka} model file, which was exported via {\sc{UFO}}~\cite{Degrande:2011ua} to {\sc{MadGraph}}~\cite{Alwall:2014hca}, in order to generate parton-level events. For the analytic results discussed later, the model file was exported to {\sc{FeynArts/FormCalc}}~\cite{Hahn:2000kx,Hahn:1998yk}, which we used to calculate the full dependence of each observable used on the operators of Eq.~\eqref{eqn:ttzops}. To generate lepton collider constraints, we minimise the $\chi^2$ between the full 5D analytic expression and SM pseudodata using {\sc{iMinuit}}, a Python/C++ implementation of the {\sc{MINUIT}}~\cite{James:1975dr} algorithm. For hadron collider constraints, we approximate the full 5D dependence of the cross-section on the operators using {\sc{Professor}}; an interpolation-based method described in Ref.~\cite{Buckley:2009bj}. We cross-check our analytic results against Monte Carlo scans using a UFO implementation of the full basis of Ref.~\cite{Grzadkowski:2010es} and find good agreement.

\begin{figure*}[t!]
\begin{center}
 \includegraphics[width=0.9\textwidth]{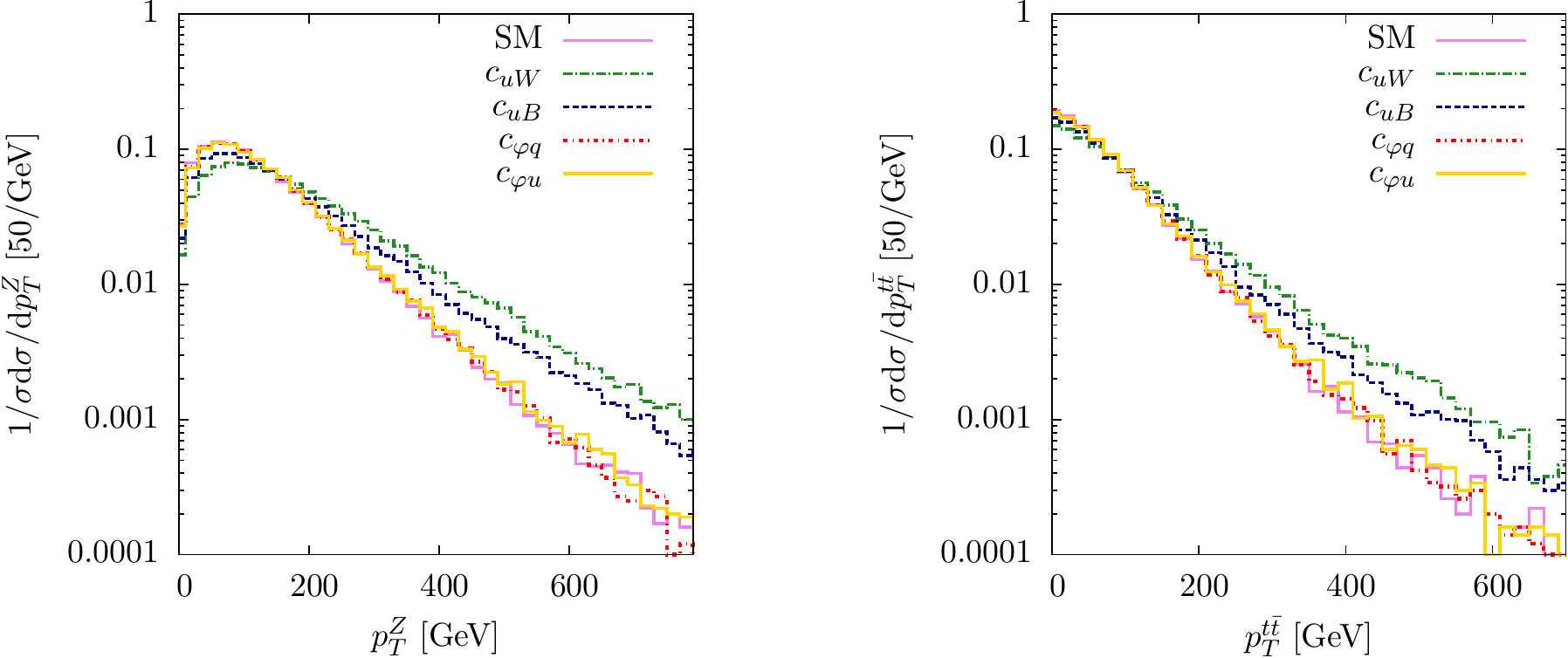}
 \caption{Kinematic distributions in $pp \to$ \ttz production at 13 \tev for the SM prediction and for the operators of Eq.~\eqref{eqn:ttzops} switched on to their maximum value allowed by current data. Left: the $Z$ boson transverse momentum spectrum. Right: $p_T^{t\bar t} = p_T^t -p_T^{\bar t}$ spectrum. All distributions are normalised to the total cross-section. Shape differences can be seen in the tails for the operators \op{uW} and \op{uB}, showing that differential distributions provide complementary information to overall rates.}
\label{fig:distributions}
 \end{center}
 \end{figure*}

\section{Hadron collider constraints}
\label{sec:had}

To appreciate the impact of the operators of Eq.~\eqref{eqn:ttzops}, in Fig.~\ref{fig:ttzxsec} we plot the ratio of the full \ttz cross-section with each operator switched on individually, to the NLO SM prediction, taken from Ref.~\cite{Maltoni:2015ena}. For ease of interpretation, we split up the cross-section into the contribution from the interference term and the quadratic term. We see firstly that the operators \op{uW} and \op{uB} have the strongest impact on the total cross-section, but this comes purely from the squared term (this was also noted in Ref.~\cite{Bylund:2016phk}). The remaining operators have a milder effect on the cross-section, but their interference term dominates. We also see that the operators \op[(3)]{\varphi q} and \op[(1)]{\varphi q} contribute the same dependence but with an opposite sign, as discussed in Sec.~\ref{sec:topew}, therefore we can only bound the linear combination \op{\varphi q}. 

The LHC bounds on the coefficients of these operators from 8 \tev \ttz production cross-sections have been presented in Refs~\cite{Buckley:2015lku}. The current constraints are weak. Since then, ATLAS and CMS have presented measurements using 13 \tev  collision data, with measured values 0.9 $\pm$ 0.3 \pb~\cite{Aaboud:2016xve} and 0.7 $\pm$ 0.21 \pb~\cite{CMS-PAS-TOP-16-017}, respectively. The constraints on the operators using these two measurements are shown in Fig.~\ref{fig:ttzconstraints}, where the coefficients are normalised to the `bar' notation \cb{i} = $\co{i}v^2/\Lambda^2$, and the operators are switched on individually. 

We see that the current constraints are still quite weak, mainly due to the large ($\sim$ 30\%) experimental uncertainties. These measurements are currently statistics dominated, so it is instructive to ask what the expected improvement is over the lifetime of the LHC. Using a constant systematic uncertainty of 10\% based on the current estimate, we also plot in Fig.~\ref{fig:ttzconstraints} the constraints using 300 \ifb and 3 \iab of SM pseudodata. We see that there will be an improvement by factors of $1.5$ to $2$ by the end of Run III, but after this the measurement is saturated by systematics.

To highlight the benefits of improving the theory description in tandem, we also show in Fig.~\ref{fig:ttzconstraints} the projected constraints if theory uncertainties are improved to 1\% from the current \ord{10\%} precision, which does not seem unreasonable over the timescales we are considering. We see again that there will be no subsequent improvement after 300~\ifb unless experimental systematics are reduced.

\begin{figure*}[t!]
\begin{center}
 \subfigure[(a)]{\includegraphics[width=0.42\textwidth]{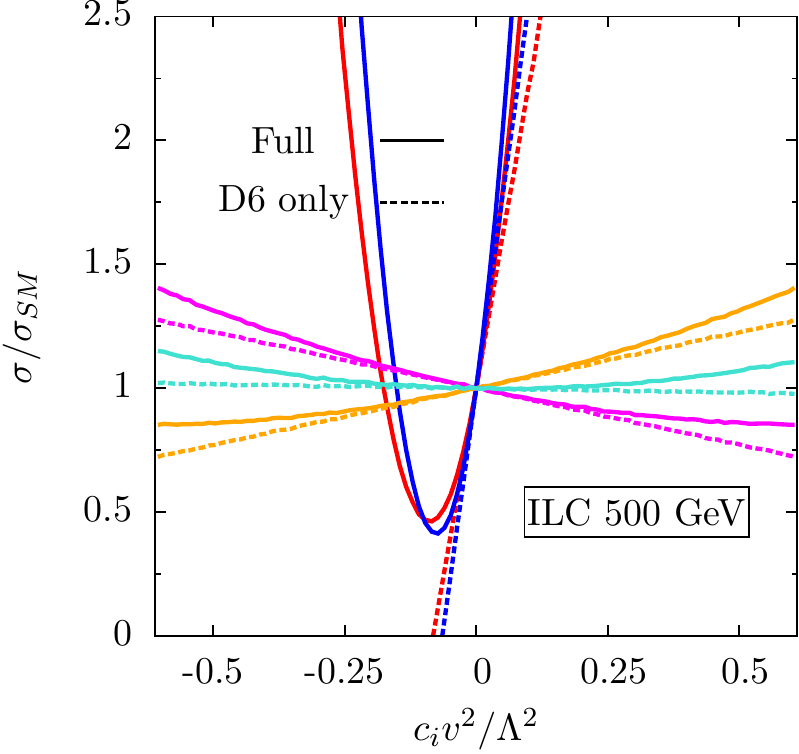}}
\hspace{0.05\textwidth}
  \subfigure[(b)]{\includegraphics[width=0.42\textwidth]{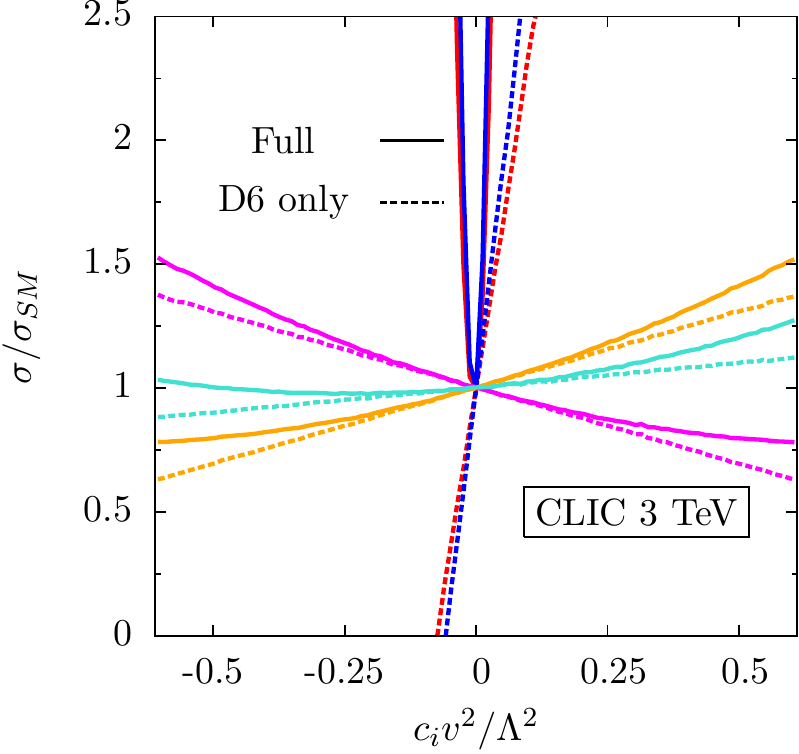}}
   \caption{(a) Ratio of the full SM \ep to \ttbar cross-section at \sqrts = 500 \gev with the operators of Eq.~\eqref{eqn:ttzops} switched on individually to the NLO Standard Model estimate. The dashed lines show the contribution from the interference term, and the solid lines show the full dependence The operator colour-coding is the same as Fig.~\ref{fig:ttzxsec}. (b) Likewise for CLIC running at \sqrts = 3 \tev. \label{fig:ilcxsec}}
  \end{center}
\end{figure*}

Finally, it should be noted that as more data becomes available, it may be possible to measure \ttz cross-sections differentially in final state quantities. Since cuts on the final state phase space can enhance sensitivity to the region where na\"ive power counting says \D6 operators become more important, differential distributions could substantially improve the fit prospects, as has already been demonstrated for \ttbar production.~\cite{Buckley:2015lku,Englert:2016aei}. 

To illustrate this, in Fig.~\ref{fig:distributions} we plot the distributions for the $Z$ boson transverse momentum and top pair transverse momentum, both for the SM only case and with each operator switched on to a value of $\cb{i} \simeq 0.3$; approximately the maximum allowed by current constraints in Fig.~\ref{fig:ttzconstraints}. We see that extra enhancement in the tail is visible for the field strength tensor operators \op{uW} and \op{uB}, due to the extra momentum dependence in the numerator from the field strength tensor. For the $\varphi$-type operators, since the interference is solely proportional to $\varphi^\dagger\varphi \to v^2/\Lambda^2$, there is no extra enhancement at high $p_T$.

We do not estimate the improvement of the fit by taking these distributions into account, since this would require proper estimates of experimental systematics and tracking the nontrivial correlations between the kinematic quantities in the massive 3-body final state. Here, we merely comment that it may be an avenue worth pursuing as more data becomes available.
  
\section{ILC constraints}
\label{sec:ilc}

Going beyond the LHC, currently, the most mature proposal is for a linear \ep collider with a centre of mass energy ranging from 250 \gev to up to 1 \tev. There are several scenarios for integrated luminosity and CM energy combinations. The most-studied is the so-called H-20 option, which involves running at 500 \gev for 500 \ifb of data, followed by 200 \ifb of data at the \ttbar threshold to perform detailed measurements of the top quark mass, and 300 \ifb of data at \sqrts = 250 \gev to maximise the machine's Higgs potential with high precision. After a luminosity upgrade, a further 3.5 \iab is gathered at \sqrts = 500 \gev, followed by another \sqrts = 250 \gev run at 1.5 \iab. Since we are most interested in the ILC mass reach for new physics, in this study we focus on the 500 \gev  ILC running\footnote{In principle, the constraints could be improved by also adding results from \sqrts = 350 \gev, however this requires more precise theory modelling to understand the impact of \D6 operators on threshold effects than our leading-order + $K$-factor analsysis.} .

\begin{figure*}[t!]
\begin{center}
 \includegraphics[width=0.9\textwidth]{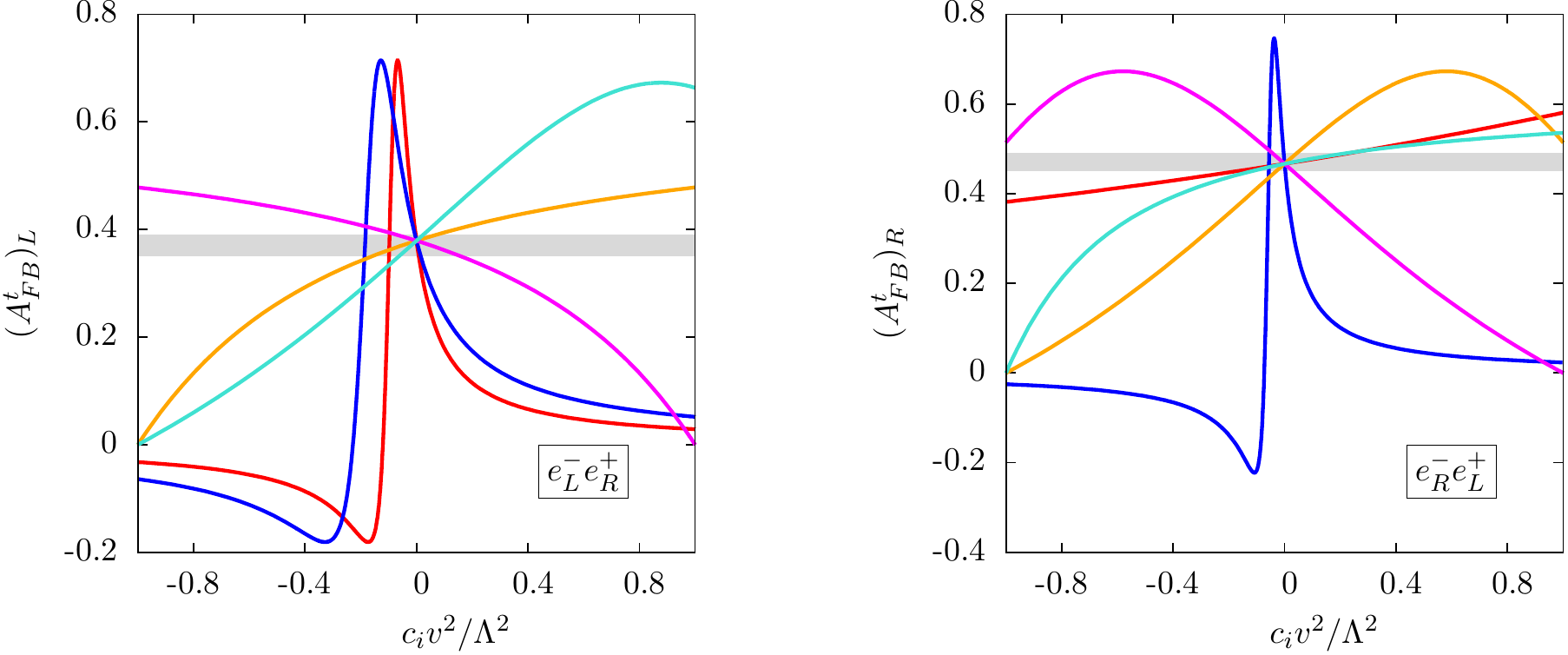}
 \caption{Full dependence of the \ttbar forward-backward asymmetry of Eq.~\eqref{eqn:asymm} on the operators  of Eq.~\eqref{eqn:ttzops} for left-handed polarised electrons (left) and right-handed polarised electrons (right) at \sqrts = 500 \gev, the operator colour coding is the same as Fig.~\ref{fig:ttzxsec}. We also show a 5\% uncertainty band around the SM prediction, to estimate the expected constraints. }
\label{fig:asymmetries}
 \end{center}
 \end{figure*}

An important parameter for lepton colliders is the energy spread of electron and positron beams (see e.g.~\cite{Boogert:2002jr}). In order to estimate the effect on our results, we use the results of \cite{Boogert:2002jr} to calculate the expected change in the cross-
section by including the effects of initial state radiation, beam spread and beamstrahlung. We find that for the typical beam profile, the associated uncertainty is not a limiting factor and we neglect these effects in the following.

\subsection{The \ttbar total cross-section}
Top pair production has a more modest rate here than at a hadron collider. The state-of-the-art Standard Model calculations for (unpolarised) \ep $\to$ \ttbar production are at N$^3$LO QCD in the narrow-width approximation~\cite{Kiyo:2009gb,Chen:2016zbz}, and at NLO QCD including off-shell effects (which have been demonstrated to be important~\cite{Nejad:2016bci}) and at NLO EW~\cite{Fleischer:2003kk} (with partial NNLO results in Ref.~\cite{Gao:2014nva}) and predict a cross-section $\sigma \simeq $ 0.57 pb. The conventional scale variation gives a QCD uncertainty at the per-mille level. While this rate is more than a factor of a thousand smaller than at the 13 \tev LHC, the process is essentially background free. Thus, after even 500 \ifb of data the statistical uncertainty will be approximately  0.2\%, and so completely subdominant to the systematics.

We can thus repeat the exercise of extracting the bounds on the coefficients of the operators of Eq.~\eqref{eqn:ttzops} using SM pseudodata. As a guide for the expected numerical constraints, we also plot the ratio of the total cross-section in the presence of the operators to the SM prediction, this time using the total (unpolarised) cross-section at the 500 \gev ILC. This is shown on the left of Fig.~\ref{fig:ilcxsec}.

We see again that the operators \op{uW} and \op{uB} are the strongest, however, unlike the case of \ttz production the interference term dominates at small $\co{i}/\Lambda^2$. The result of this is that there is a cancellation between the interference and quadratic terms at approximately $\co{i}/\Lambda^2 \simeq -3~\tev^{-2}$, leading to a SM-like cross-section and a second, degenerate minimum in the $\chi^2$. The constraints obtained from a one-at-a-time fit of these operators to the 
SM pseudodata is shown in the red bars on the right of Fig.~\ref{fig:ilcglobal}. 

The operators \op{uW} and \op{uB} are very tightly constrained, due to their much stronger impact on the cross-section stemming from the extra momentum dependence flowing through the vertex. The $\varphi$-type operators are more weakly constrained, but on the whole the constraints are typically 10 to 100 times stronger than for the LHC \ttz production projections in Sec.~\ref{sec:had}, which is unsurprising giving the difference in precision. 

\begin{figure*}[t!]
\begin{center}
 \subfigure[(a)]{\includegraphics[width=0.475\textwidth]{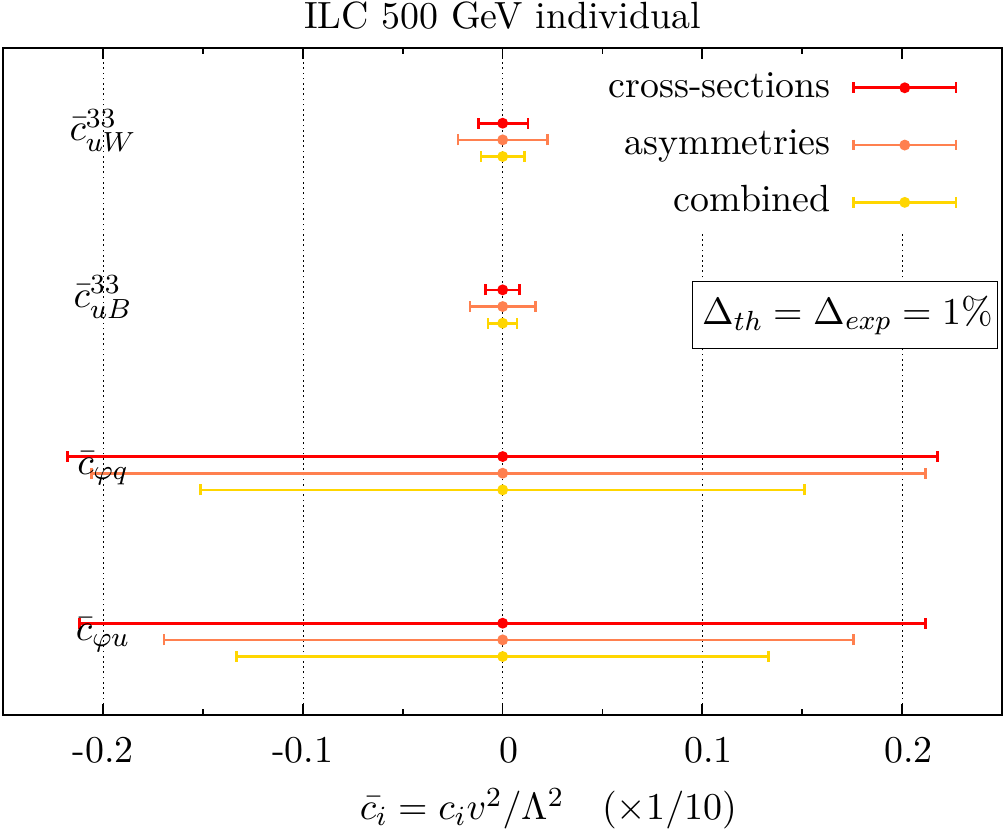}}\hfill
  \subfigure[(b)]{\includegraphics[width=0.475\textwidth]{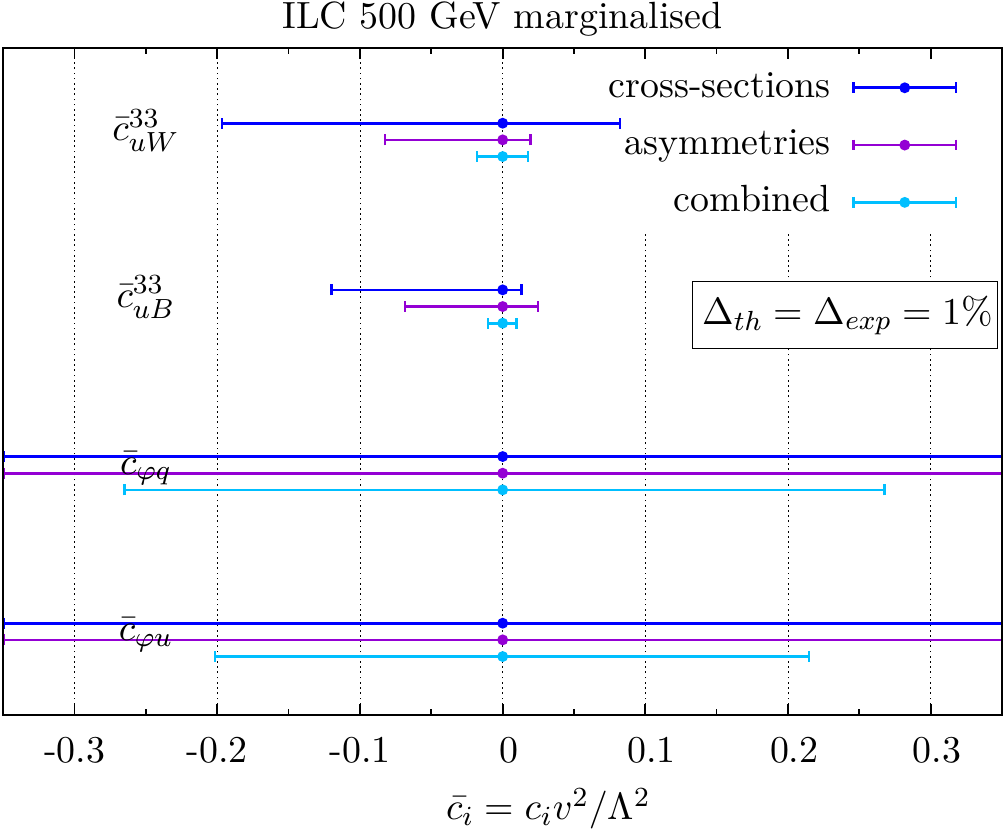}}
   \caption{95\% confidence ranges for the operators we consider here, from the 500 \gev ILC, assuming 1\% theoretical and experimental uncertainties, by fitting to cross-sections, asymmetries, and the combination, with each operator considered individually (a) or in a 5D fit (b). To display both on the same plot, we scale the individual constraints up by a factor of 10, so that the bottom axis is actually \cb{i}/10.\label{fig:ilcglobal}}
  \end{center}
\end{figure*}

Individual constraints are less useful in practice, however. Firstly, in a plausible UV scenario that would generate these operators one would typically expect more than one to be generated at once, so that one-at-a-time constraints cannot be straightforwardly linked to a specific `top-down' model. Secondly, there can in general be cancellations between different operators for a given observable that can yield spurious local minima and disrupt the fit. This would not be visible in the individual constraints, and so would obscure degeneracies in the operator set that could be broken by considering different observables. Therefore, we also consider constraints where we marginalise over the remaining three coefficients in the fit. These are shown in the blue bars on the right of Fig.~\ref{fig:ilcglobal}.\footnote{Note, however, that a full marginalisation will be overly conservative when confronting a concrete UV model.}

We see that, with the exception of \op{uW} and \op{uB}, marginalising over the full operator set wipes out the constraints. This is because even for large values of coefficients, the pull that a particular operator has on the cross-section can easily be cancelled by another operator. We can conclude that, despite the impressive precision that can be achieved in extracting the cross-section, it has limited use in constraining new physics in a simultaneous global fit of several operators. It is worthwhile to make use of other measurements.

\subsection{Polarised beams}
One of the principal strengths of lepton colliders is that the polarisation of the incoming beams can be finely controlled, so that the relative contributions between different subprocesses to a given final state can be tuned. Moreover, because the dependence of top observables on the operators of Eq.~\eqref{eqn:ttzops} depends strongly on the initial state polarisation, varying the settings increases the number of independent measurements that can be used to place bounds in a global fit.

To emphasise this point, we study the forward-backward asymmetry, defined as
\begin{equation}
A^t_{FB} = \frac{N(\cos\theta_t > 0)-N(\cos\theta_t < 0)}{N(\cos\theta_t > 0)+N(\cos\theta_t < 0)},
\label{eqn:asymm}
\end{equation}
where $\theta_t$ is the polar angle between the top quark and the incoming electron, for three incoming beam polarisation settings: unpolarised beams, denoted $(A^t_{FB})_U$; a fully left-handed initial polarised electron beam and fully right-handed polarised positron beam, denoted $(A^t_{FB})_L$; and vice versa, denoted $(A^t_{FB})_R$. The SM predictions for these settings at tree level are $\{(A^t_{FB})_U,(A^t_{FB})_L,(A^t_{FB})_R\} \simeq \{0.40, 0.37, 0.47\}$, which agree well with the full NNLO QCD estimates~\cite{Bernreuther:2006vp,Gao:2014eea}. The dependence of these asymmetries on the operators of Eq.~\eqref{eqn:ttzops} is shown in Fig.~\ref{fig:asymmetries}.

We see that the dependence on the operators distinctively depends on the initial state polarisations. For the $(A^t_{FB})_L$ case, we again see the large interference-square cancellation in the gauge-type operators \op{uW} and \op{uB}. For the right-handed case the impact of \op{uW} is much milder. For both cases we see that the operators \op[(3)]{\varphi q} and \op[(1)]{\varphi q} pull the prediction in opposite directions. Most encouragingly, we see that the departure from the SM prediction is now much stronger for the $\varphi$-type operators than the total cross-section, which should lead to a sizeable improvement in the final constraints.

To generate these constraints, we consider a global fit of the four operators to six observables: 
\begin{equation}
\{(A^t_{FB})_U,(A^t_{FB})_L,(A^t_{FB})_R, (\sigma^{\ttbar}_{\text{tot}})_U, (\sigma^{\ttbar}_{\text{tot}})_L, (\sigma^{\ttbar}_{\text{tot}})_R \}. 
\end{equation}

In extracting the constraints, we consider the more realistic ILC polarisation capabilities $\mathcal{P}_{e^-} = \pm$ 0.8, $\mathcal{P}_{e^+} = \mp$ 0.3, noting that the cross-section for arbitrary \ep polarisations is related to the fully polarised one by~\cite{Hikasa:1985qi,MoortgatPick:2005cw}
\begin{equation}
\begin{split}
\sigma_{\mathcal{P}_{e^-}\mathcal{P}_{e^+}} = &\frac{1}{4} \{ (1+\mathcal{P}_{e^-})(1-\mathcal{P}_{e^+})\sigma_{\text{RL}} \\
								      &+ (1-\mathcal{P}_{e^-})(1+\mathcal{P}_{e^+})\sigma_{\text{LR}} \} ,
\end{split}								      
\end{equation}

\begin{figure*}[t!]
\begin{center}
 \subfigure[(a)]{\includegraphics[width=0.475\textwidth]{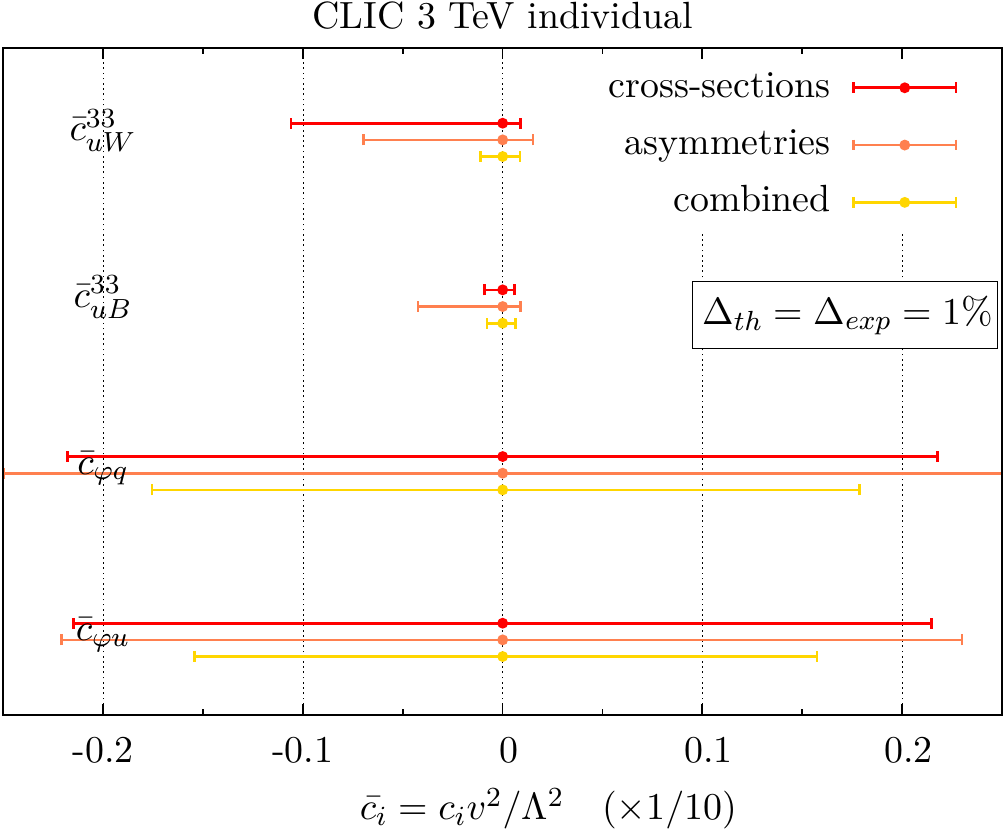}}\hfill
  \subfigure[(b)]{\includegraphics[width=0.475\textwidth]{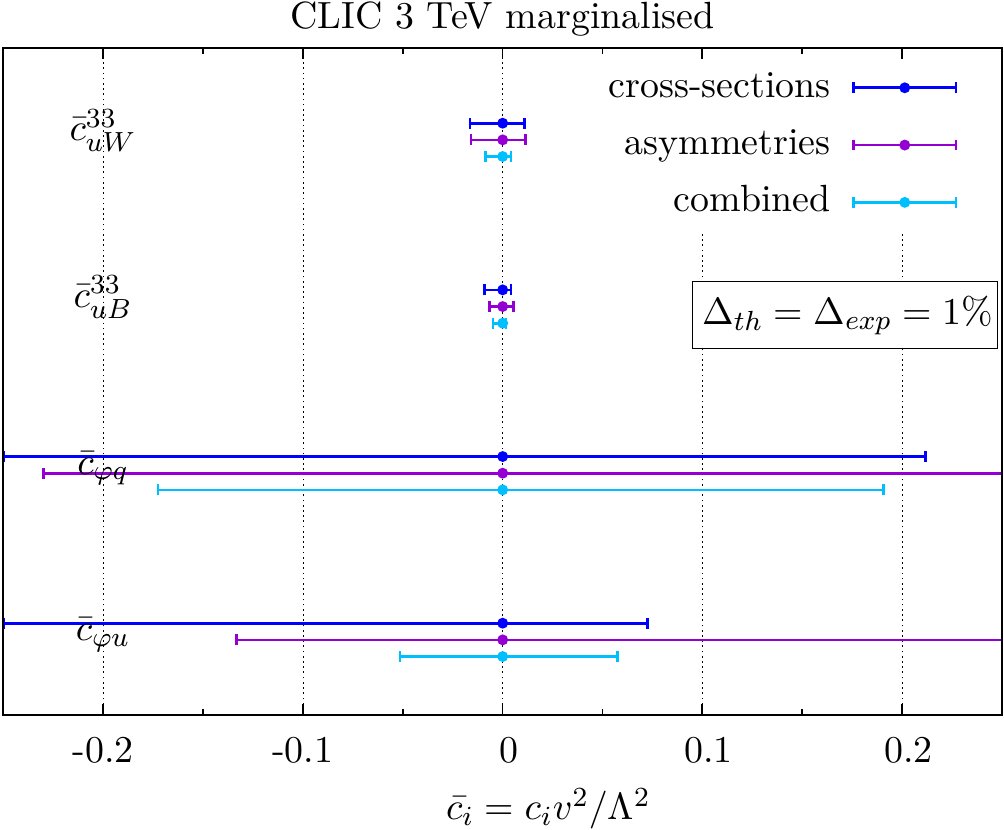}}
   \caption{95\% confidence ranges for the operators we consider here, from CLIC running at \sqrts = 3 \tev, assuming 1\% theoretical and experimental uncertainties, by fitting to cross-sections, asymmetries, and the combination, with each operator considered individually (a) or in a 5D fit (b). To display both on the same plot, we scale the individual constraints up by a factor of 10, so that the bottom axis is actually \cb{i}/10.\label{fig:clicglobal}}
  \end{center}
\end{figure*}

where $\sigma_{\text{RL}}$ is the cross-section for fully right-handed polarised electrons and fully left-handed polarised positrons and $\sigma_{\text{RL}}$ is vice versa (the $\sigma_{\text{RR}}$ and $\sigma_{\text{LL}}$ components vanish for $p$-wave annihilation into spin-1 bosons). Performing a $\chi^2$ fit of the full analytic expression for each observable, using SM pseudodata with 1\% experimental error bars (based on studies in Refs.~\cite{Amjad:2013tlv,Amjad:2015mma}) and SM theory uncertainties of 1\% (based on the calculations of Refs.~\cite{Kiyo:2009gb,Chen:2016zbz,Bernreuther:2006vp,Fleischer:2003kk,Gao:2014eea}) the individual and marginalised constraints on these operators are shown in Fig.~\ref{fig:ilcglobal}.

At the level of individual operators, the constraints are not improved drastically by adding in asymmetry information. For the global fit, however, the constraints lead to much stronger bounds than for fitting to cross-sections (although the marginalisation typically weakens the overall constraints by a factor \ord{10}).

We see that the constraints are again much stronger for the field strength operators \op{uW} and \op{uB}, where the constraints are at the $|\cb{i}|  \lesssim 10^{-4}$ level for the individual constraints and $|\cb{i}|  \lesssim 10^{-2}$ for the marginalised case, corresponding to a mass reach of $\Lambda \gtrsim 10$ \tev and $\Lambda \gtrsim 2.16$ \tev, respectively, assuming $\co{i} \simeq 1 $. The weakest constraints are on the operators \op[(3)]{\varphi q} (\op[(1)]{\varphi q}), which translate into bounds on $\Lambda$ of roughly 700 \gev. 

While it is encouraging that the bounds are consistent with an EFT formulation, in the sense that $\Lambda \gg \sqrt{s}$, the ILC mass reach for the scale of new physics that would generate these operators is still low. We note, however, that these bounds are on the conservative side, since other observables such as oblique parameters and LEP asymmetries contribute complementary information that will in general tighten them. To keep this fit self-contained, we postpone this discussion until Sec.~\ref{sec:lep}.

\section{CLIC constraints}
\label{sec:clic}

The Compact Linear Collider (CLIC) project~\cite{CLIC:2016zwp,Abramowicz:2016zbo}, with its larger maximum centre-of-mass energy \sqrts = 3 \tev, will be in a stronger position to discover the effects of some higher-dimensional operators, whose effects na\"ively scale with the CM energy as $s/\Lambda^2$. There are two main running scenarios, but both envisage total integrated luminosities of 500 \ifb at \sqrts = 500 \gev, 1.5 \iab at 1.4 or 1.5 \tev, and 2 \iab at 3 \tev. Again, we focus on the highest energy setting \sqrts = 3 \tev, to maximise discovery potential for non-resonant new physics through \D6 operators.

Moving further away from the \ttbar threshold, the total \ep $\to Z^*/\gamma \to$ \ttbar rate is smaller than at the ILC; at \sqrts = 3 \tev it is around 20 \fb, which means for the total forecast integrated luminosity at this energy  there will be a statistical uncertainty of $\simeq$ 0.5\%. A total experimental uncertainty of 1\% may therefore be too optimistic an estimate once systematics are fully itemised. Nonetheless, for ease of comparison with the ILC figures, we take this as a baseline, and the corresponding constraints, using the same observables and beam settings, are shown in Fig.~\ref{fig:clicglobal}.

We see that for the individual fit, CLIC constraints are of the same order of magnitude as the ILC ones.\footnote{This is in contrast to Higgs sector constraints from \ep $\to hZ$, where the projected sensitivity is extremely dependent on the momentum flow through the vertex, leading to better overall CLIC constraints~\cite{Ellis:2017kfi}.} Although the direct sensitivity to the operators is enhanced, we see that as we move away from the \ttbar threshold, the interference effect of the $\varphi$-type operators is much smaller. This is not the case for the operators \op{uW} and \op{uB}, whose contributions stem mainly from the $(\D6)^2$ term, as seen on the right of Fig.~\ref{fig:ilcxsec}, which receives no suppression. Their individual constraints are close to the ILC values, indicating that energy scale is not the dominant factor driving these limits, but rather the theory and experimental uncertainties which saturate the sensitivity, which we do not vary.

For the more general marginalised fit, we see again that combining cross-section and asymmetry measurements will break blind directions in the fit, leading to much more powerful overall constraints. Unlike for the case of the ILC, however, care must be taken in interpreting these limits in terms of the mass scale of a particular UV model. The marginalised constraint $|\cb{\varphi u}| \lesssim 0.05$, for example, corresponds to a mass scale $\Lambda/\sqrt{c} \gtrsim $ 1.1 \tev, which is less than the energy scale probed in the interaction, so that the constraint can only be linked to a particular model if it is very weakly coupled: $g_* \ll 1$. 

\begin{figure}[t!]
\begin{center}
 \includegraphics[width=0.22\textwidth]{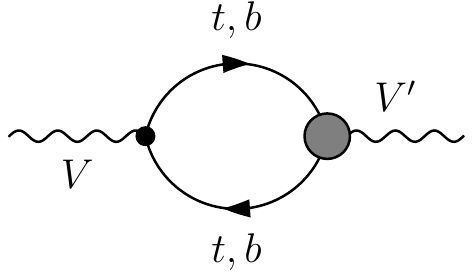}
  \caption{\label{fig:pols} Representative Feynman diagram contributing to the $S,T,U$ parameters at one-loop. The grey-shaded area marks a possible dimension six insertion while the black dot represents a SM vertex of the $V-V'$ polarisation function, $V,V'=W^\pm, Z,\gamma$.}
\end{center}
\end{figure}

\section{Beyond $e^+e^-\to t\bar t$: Precision Observables}
\label{sec:lep}
Obviously the direct constraints that we have focused on in this work do not exist in a vacuum and the interplay of direct and indirect sensitivity plays an important part in ultimately obtaining the best constraints for a given model (see \cite{Berthier:2015oma,Ghezzi:2015vva}). To put the expected direct constraints detailed above into perspective we analyse the impact of the considered operators on LEP precision observables. Note, that these $Z$ resonance observables are sensitive to a plethora of other new interactions and a direct comparison is not immediately straightforward~\cite{Berthier:2015oma}. Nonetheless, there is significant discriminative power that is worthwhile pointing out, which we will discuss in the following.

\begin{figure}[t!]
\begin{center}
 \includegraphics[width=0.48\textwidth]{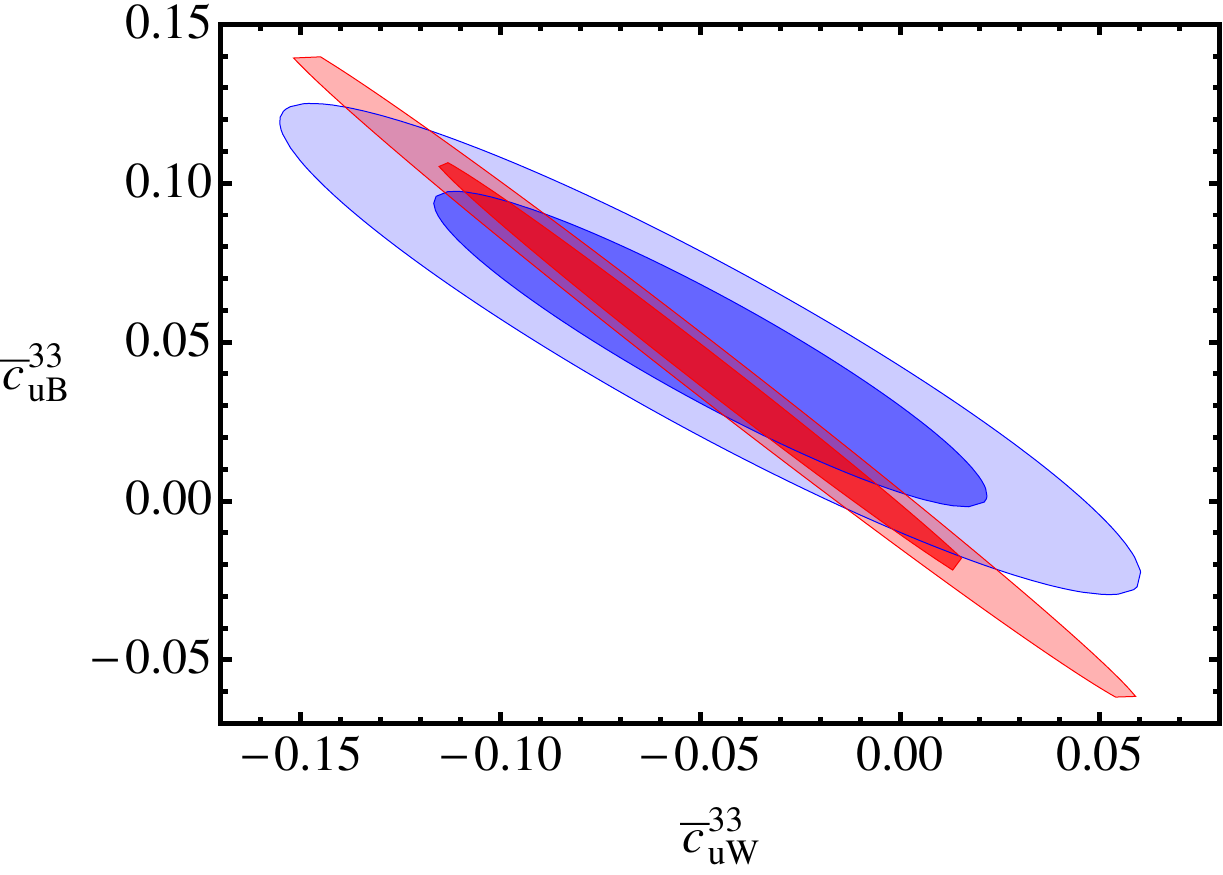}
  \caption{\label{fig:stu} Contour of the $S,T,U$ fit reported in \cite{Baak:2014ora} for specifically the operators $\op{uW},\op{uB}$, which are unconstrained by down-sector measurements. All other Wilson coefficients are chosen to be zero. The dark and light shaded areas represent 68\% and 95\% confidence levels for this projections, while the blue contour uses $\mu_R=m_Z$ and the red contour $\mu_R=1~\text{TeV}$.}
\end{center}
\end{figure}

\subsection{Oblique corrections}
The $S,T,U$ parameters~\cite{Peskin:1990zt,Peskin:1991sw} (see also \cite{Ross:1975fq}) are standard observables that capture oblique deviations in the SM electroweak gauge sector from the SM paradigm~\cite{Grojean:2013kd,Barbieri:2004qk} through modifications of the gauge boson two-point functions. The operators considered in this work modify these at the one-loop level through diagrams of the type shown in Fig.~\ref{fig:pols}. Throughout we perform our calculation in dimensional regularisation.

The definitions of $S,T,U$, see~\cite{Peskin:1990zt,Peskin:1991sw,Barbieri:2004qk}, are such that in the SM all divergencies cancel when replacing the renormalised polarisation functions by their bare counterparts. The modifications of Fig.~\ref{fig:pols}, however, induce additional divergencies due to the dimension 6 parts and the introduction of two-point function counterterms is essential to obtain a UV-finite result, see also~\cite{Ghezzi:2015vva}. This leads to a regularisation scale $\mu_R$ dependence of the dimension 6 amplitude parts after renormalisation. It is this part which we focus on as we choose the SM with a 125 GeV Higgs as reference point~\cite{Baak:2014ora}.

\begin{figure*}[t!]
\begin{center}
 \parbox{0.47\textwidth}{\centering
 \includegraphics[height=0.31\textwidth]{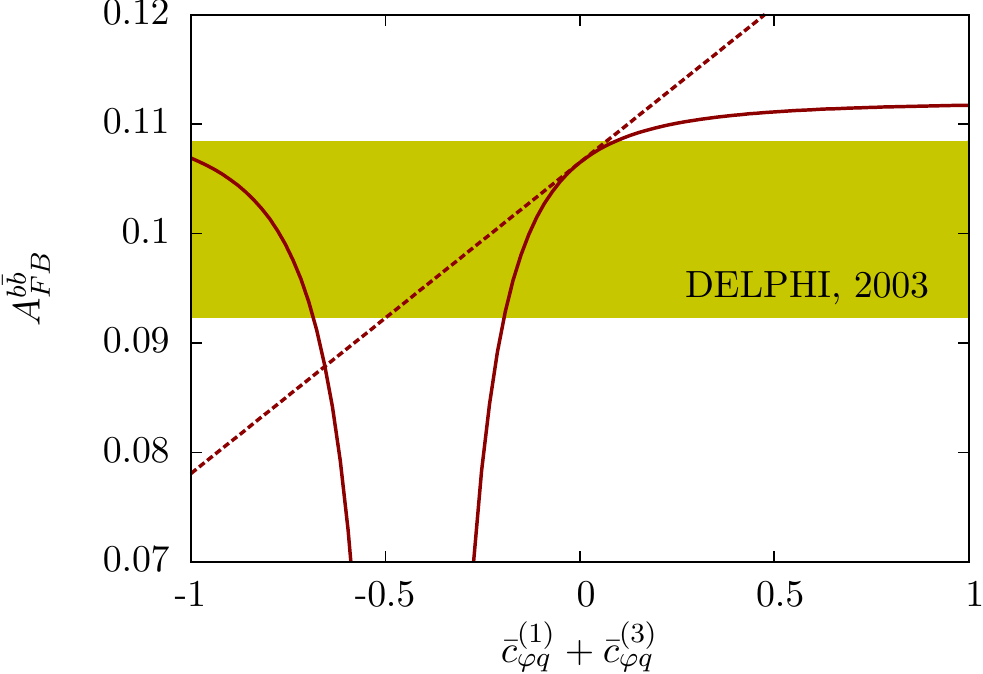}
  \caption{\label{fig:afbbb} Forward-backward asymmetry (linearised, dashed, and full results, solid) as a function of $\co[(3)]{\varphi q}+ \co[(1)]{\varphi q}$. The exclusion contour is taken from DELPHI collaboration's Ref.~\cite{Abdallah:2003gp} for the most constraining measurement at $\sqrt{s}=91.26~\text{GeV}$.}}
  \hspace{0.6cm}
 \parbox{0.47\textwidth}{
 \vspace{-0.2cm}
 \centering
 \includegraphics[height=0.31\textwidth]{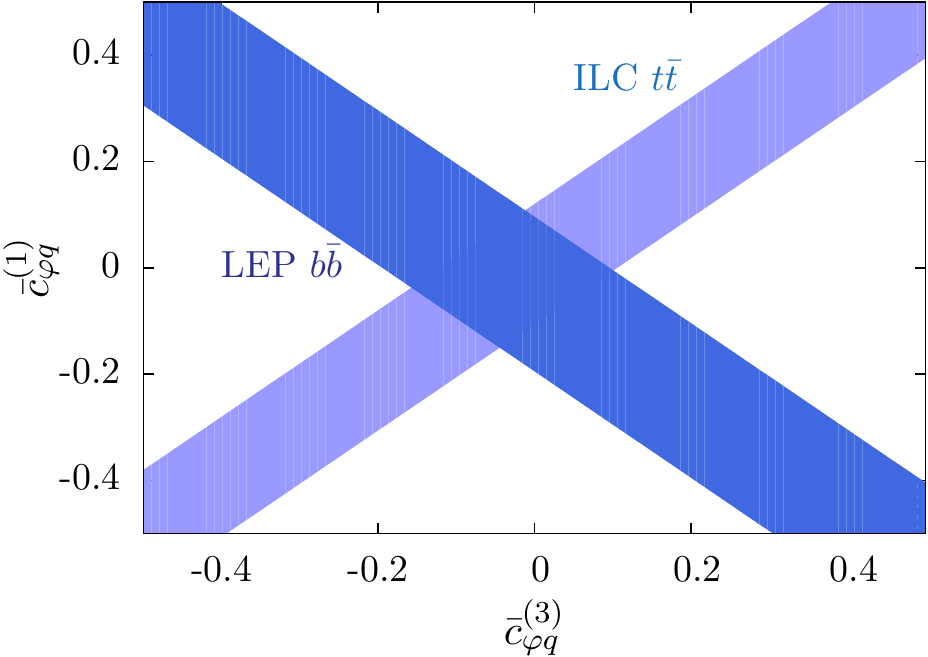}
  \caption{\label{fig:lepilc} Allowed 95\% confidence regions for the Wilson coefficients \cb[(3)]{\varphi u} and \cb[(1)]{\varphi u} obtained from combining the information from ILC \ttbar asymmetries and cross-sections (dark blue) and LEP1 $b\bar b$ measurements (lighter blue).}}
\end{center}
\end{figure*}

As can be seen from Fig.~\ref{fig:stu}, oblique corrections constrain the range of the considered parameter (setting all other contributions to zero). Given that we deal with an effective theory, there is a dependence on the UV cut-off $\mu_R$ and constraints can be sharpened by requiring that cut-off to lie well above the TeV scale. As such, the oblique corrections are explicitly sensitive to concrete UV completions, and care needs to be taken not to over-emphasise their impact.

\subsection{Non-oblique corrections}
A well-measured quantity at LEP is the $Zb \bar b$ vertex, which enters the prediction of the bottom forward-backward asymmetry $A_{FB}^{b\bar b}$, see e.g.~\cite{Abdallah:2003gp}. Similar to the operators in Eq.~\eqref{eqn:ttzops}, in the generic dimension six approach we can expect similar operators for the down-sector of the 3rd fermion family. These will modify the interactions along the same lines as we focused on above for the top sector. However, due to the different isospin properties, the bottom forward backward asymmetry is now sensitive to the sum $\co[(3)]{\varphi q}+\co[(1)]{\varphi q}$. This leads to a complementary constraint by the LEP forward backward asymmetry compared to the direct measurements in $t\bar t$, Fig.~\ref{fig:afbbb}.

Moreover, the constraints on \co[(3)]{\varphi q}+\co[(1)]{\varphi q} from $A_{FB}^{b\bar b}$ can be combined with the constraints on  \co[(3)]{\varphi q}$-$\co[(1)]{\varphi q} to extract independent bounds on \co[(3)]{\varphi q} and \co[(1)]{\varphi q}. This is shown in Fig.~\ref{fig:lepilc}. Care should be taken when interpreting these constraints individually, however. We are considering marginalised bounds for the ILC constraint but only one operator combination for the LEP bound. In general, other operators that we do not consider here will impact the $Zb\bar b$ vertex at tree level and in general weaken the bound. This serves as a useful visualisation, however, of the complementarity between past and future colliders in constraining these operators.

\section{Summary and conclusions}
\label{sec:conc}
Given the unsatisfactory precision of current probes of top quark electroweak couplings from hadron collider measurements, they must be a key priority in the physics agenda of any future linear \ep collider. By parameterising non-standard top couplings through \D6 operators, we have analysed the potential for the ILC and CLIC to improve the current precision of the top electroweak sector. Unsurprisingly, if experimental precision would match current estimates, and theory uncertainties can be brought to the same level, the current constraints can be drastically improved by both colliders, with associated bounds on the scale of new physics typically in the 1 \tev to few \tev range, depending on the assumed coupling structures of the underlying model. Using asymmetry measurements as well as cross-sections will be crucial to this endeavour, as will collecting large datasets with several incoming beam polarisations. 

We have found that, unlike for the Higgs sector, the large increase in centre-of-mass energy at CLIC does not necessarily offer a competitive advantage over the ILC for bounding new top interactions by the operators we consider, and bounds on the operators that we consider are typically stronger at the latter, though in simultaneous 4D fits the difference is not striking. For some of the operators we consider, the bounds derived from CLIC fits correspond to mass scales smaller than the CM energy that we consider, which can call into question the validity of the EFT description, unless the CLIC sensitivity can exceed the expectations we forecast here.

By combining $Z$-pole measurements from LEP1 with \ttbar measurements (and future improved electroweak precision measurements), one can in principle break degeneracies in the operator set and disentangle individual operators that could previously be only bounded in combinations. We showed this for the LEP forward-backward asymmetry, this could be improved by fitting other precision electroweak observables too. Care must be taken in interpreting the associated constraints, however, as both sets of measurements will in general talk to other operators for which there is no complementarity, and a more systematic approach taking into account EFT loop corrections would have be undertaken before these numerical bounds can be taken at face value.

\acknowledgements
MR is supported in part by the UK Science and Technology Facilities Council (STFC) under grant ST/L000446/1.

\bibliography{bibfile}

\begin{thebibliography}{59}
\expandafter\ifx\csname natexlab\endcsname\relax\def\natexlab#1{#1}\fi
\expandafter\ifx\csname bibnamefont\endcsname\relax
  \def\bibnamefont#1{#1}\fi
\expandafter\ifx\csname bibfnamefont\endcsname\relax
  \def\bibfnamefont#1{#1}\fi
\expandafter\ifx\csname citenamefont\endcsname\relax
  \def\citenamefont#1{#1}\fi
\expandafter\ifx\csname url\endcsname\relax
  \def\url#1{\texttt{#1}}\fi
\expandafter\ifx\csname urlprefix\endcsname\relax\def\urlprefix{URL }\fi
\providecommand{\bibinfo}[2]{#2}
\providecommand{\eprint}[2][]{\url{#2}}

\bibitem[{\citenamefont{Georgi et~al.}(1995)\citenamefont{Georgi, Kaplan,
  Morin, and Schenk}}]{Georgi:1994ha}
\bibinfo{author}{\bibfnamefont{H.}~\bibnamefont{Georgi}},
  \bibinfo{author}{\bibfnamefont{L.}~\bibnamefont{Kaplan}},
  \bibinfo{author}{\bibfnamefont{D.}~\bibnamefont{Morin}}, \bibnamefont{and}
  \bibinfo{author}{\bibfnamefont{A.}~\bibnamefont{Schenk}},
  \bibinfo{journal}{Phys. Rev.} \textbf{\bibinfo{volume}{D51}},
  \bibinfo{pages}{3888} (\bibinfo{year}{1995}), \eprint{hep-ph/9410307}.

\bibitem[{\citenamefont{Contino et~al.}(2003)\citenamefont{Contino, Nomura, and
  Pomarol}}]{Contino:2003ve}
\bibinfo{author}{\bibfnamefont{R.}~\bibnamefont{Contino}},
  \bibinfo{author}{\bibfnamefont{Y.}~\bibnamefont{Nomura}}, \bibnamefont{and}
  \bibinfo{author}{\bibfnamefont{A.}~\bibnamefont{Pomarol}},
  \bibinfo{journal}{Nucl. Phys.} \textbf{\bibinfo{volume}{B671}},
  \bibinfo{pages}{148} (\bibinfo{year}{2003}), \eprint{hep-ph/0306259}.

\bibitem[{\citenamefont{Agashe et~al.}(2005)\citenamefont{Agashe, Contino, and
  Pomarol}}]{Agashe:2004rs}
\bibinfo{author}{\bibfnamefont{K.}~\bibnamefont{Agashe}},
  \bibinfo{author}{\bibfnamefont{R.}~\bibnamefont{Contino}}, \bibnamefont{and}
  \bibinfo{author}{\bibfnamefont{A.}~\bibnamefont{Pomarol}},
  \bibinfo{journal}{Nucl. Phys.} \textbf{\bibinfo{volume}{B719}},
  \bibinfo{pages}{165} (\bibinfo{year}{2005}), \eprint{hep-ph/0412089}.

\bibitem[{\citenamefont{Nilles}(1984)}]{Nilles:1983ge}
\bibinfo{author}{\bibfnamefont{H.~P.} \bibnamefont{Nilles}},
  \bibinfo{journal}{Phys. Rept.} \textbf{\bibinfo{volume}{110}},
  \bibinfo{pages}{1} (\bibinfo{year}{1984}).

\bibitem[{\citenamefont{Contino et~al.}(2016)\citenamefont{Contino, Falkowski,
  Goertz, Grojean, and Riva}}]{Contino:2016jqw}
\bibinfo{author}{\bibfnamefont{R.}~\bibnamefont{Contino}},
  \bibinfo{author}{\bibfnamefont{A.}~\bibnamefont{Falkowski}},
  \bibinfo{author}{\bibfnamefont{F.}~\bibnamefont{Goertz}},
  \bibinfo{author}{\bibfnamefont{C.}~\bibnamefont{Grojean}}, \bibnamefont{and}
  \bibinfo{author}{\bibfnamefont{F.}~\bibnamefont{Riva}},
  \bibinfo{journal}{JHEP} \textbf{\bibinfo{volume}{07}}, \bibinfo{pages}{144}
  (\bibinfo{year}{2016}), \eprint{1604.06444}.

\bibitem[{\citenamefont{Buckley et~al.}(2015)\citenamefont{Buckley, Englert,
  Ferrando, Miller, Moore, Russell, and White}}]{Buckley:2015nca}
\bibinfo{author}{\bibfnamefont{A.}~\bibnamefont{Buckley}},
  \bibinfo{author}{\bibfnamefont{C.}~\bibnamefont{Englert}},
  \bibinfo{author}{\bibfnamefont{J.}~\bibnamefont{Ferrando}},
  \bibinfo{author}{\bibfnamefont{D.~J.} \bibnamefont{Miller}},
  \bibinfo{author}{\bibfnamefont{L.}~\bibnamefont{Moore}},
  \bibinfo{author}{\bibfnamefont{M.}~\bibnamefont{Russell}}, \bibnamefont{and}
  \bibinfo{author}{\bibfnamefont{C.~D.} \bibnamefont{White}},
  \bibinfo{journal}{Phys. Rev.} \textbf{\bibinfo{volume}{D92}},
  \bibinfo{pages}{091501} (\bibinfo{year}{2015}), \eprint{1506.08845}.

\bibitem[{\citenamefont{Buckley et~al.}(2016)\citenamefont{Buckley, Englert,
  Ferrando, Miller, Moore, Russell, and White}}]{Buckley:2015lku}
\bibinfo{author}{\bibfnamefont{A.}~\bibnamefont{Buckley}},
  \bibinfo{author}{\bibfnamefont{C.}~\bibnamefont{Englert}},
  \bibinfo{author}{\bibfnamefont{J.}~\bibnamefont{Ferrando}},
  \bibinfo{author}{\bibfnamefont{D.~J.} \bibnamefont{Miller}},
  \bibinfo{author}{\bibfnamefont{L.}~\bibnamefont{Moore}},
  \bibinfo{author}{\bibfnamefont{M.}~\bibnamefont{Russell}}, \bibnamefont{and}
  \bibinfo{author}{\bibfnamefont{C.~D.} \bibnamefont{White}},
  \bibinfo{journal}{JHEP} \textbf{\bibinfo{volume}{04}}, \bibinfo{pages}{015}
  (\bibinfo{year}{2016}), \eprint{1512.03360}.

\bibitem[{\citenamefont{Bessidskaia~Bylund
  et~al.}(2016)\citenamefont{Bessidskaia~Bylund, Maltoni, Tsinikos, Vryonidou,
  and Zhang}}]{Bylund:2016phk}
\bibinfo{author}{\bibfnamefont{O.}~\bibnamefont{Bessidskaia~Bylund}},
  \bibinfo{author}{\bibfnamefont{F.}~\bibnamefont{Maltoni}},
  \bibinfo{author}{\bibfnamefont{I.}~\bibnamefont{Tsinikos}},
  \bibinfo{author}{\bibfnamefont{E.}~\bibnamefont{Vryonidou}},
  \bibnamefont{and} \bibinfo{author}{\bibfnamefont{C.}~\bibnamefont{Zhang}},
  \bibinfo{journal}{JHEP} \textbf{\bibinfo{volume}{05}}, \bibinfo{pages}{052}
  (\bibinfo{year}{2016}), \eprint{1601.08193}.

\bibitem[{\citenamefont{Durieux et~al.}(2015)\citenamefont{Durieux, Maltoni,
  and Zhang}}]{Durieux:2014xla}
\bibinfo{author}{\bibfnamefont{G.}~\bibnamefont{Durieux}},
  \bibinfo{author}{\bibfnamefont{F.}~\bibnamefont{Maltoni}}, \bibnamefont{and}
  \bibinfo{author}{\bibfnamefont{C.}~\bibnamefont{Zhang}},
  \bibinfo{journal}{Phys. Rev.} \textbf{\bibinfo{volume}{D91}},
  \bibinfo{pages}{074017} (\bibinfo{year}{2015}), \eprint{1412.7166}.

\bibitem[{\citenamefont{Röntsch and Schulze}(2014)}]{Rontsch:2014cca}
\bibinfo{author}{\bibfnamefont{R.}~\bibnamefont{Röntsch}} \bibnamefont{and}
  \bibinfo{author}{\bibfnamefont{M.}~\bibnamefont{Schulze}},
  \bibinfo{journal}{JHEP} \textbf{\bibinfo{volume}{07}}, \bibinfo{pages}{091}
  (\bibinfo{year}{2014}), \bibinfo{note}{[Erratum: JHEP09,132(2015)]},
  \eprint{1404.1005}.

\bibitem[{\citenamefont{Rosello and Vos}(2016)}]{Rosello:2015sck}
\bibinfo{author}{\bibfnamefont{M.~P.} \bibnamefont{Rosello}} \bibnamefont{and}
  \bibinfo{author}{\bibfnamefont{M.}~\bibnamefont{Vos}}, \bibinfo{journal}{Eur.
  Phys. J.} \textbf{\bibinfo{volume}{C76}}, \bibinfo{pages}{200}
  (\bibinfo{year}{2016}), \eprint{1512.07542}.

\bibitem[{\citenamefont{Schulze and Soreq}(2016)}]{Schulze:2016qas}
\bibinfo{author}{\bibfnamefont{M.}~\bibnamefont{Schulze}} \bibnamefont{and}
  \bibinfo{author}{\bibfnamefont{Y.}~\bibnamefont{Soreq}},
  \bibinfo{journal}{Eur. Phys. J.} \textbf{\bibinfo{volume}{C76}},
  \bibinfo{pages}{466} (\bibinfo{year}{2016}), \eprint{1603.08911}.

\bibitem[{\citenamefont{de~Blas et~al.}(2015)\citenamefont{de~Blas, Chala, and
  Santiago}}]{deBlas:2015aea}
\bibinfo{author}{\bibfnamefont{J.}~\bibnamefont{de~Blas}},
  \bibinfo{author}{\bibfnamefont{M.}~\bibnamefont{Chala}}, \bibnamefont{and}
  \bibinfo{author}{\bibfnamefont{J.}~\bibnamefont{Santiago}},
  \bibinfo{journal}{JHEP} \textbf{\bibinfo{volume}{09}}, \bibinfo{pages}{189}
  (\bibinfo{year}{2015}), \eprint{1507.00757}.

\bibitem[{\citenamefont{Aguilar-Saavedra
  et~al.}(2007)\citenamefont{Aguilar-Saavedra, Carvalho, Castro, Veloso, and
  Onofre}}]{AguilarSaavedra:2006fy}
\bibinfo{author}{\bibfnamefont{J.~A.} \bibnamefont{Aguilar-Saavedra}},
  \bibinfo{author}{\bibfnamefont{J.}~\bibnamefont{Carvalho}},
  \bibinfo{author}{\bibfnamefont{N.~F.} \bibnamefont{Castro}},
  \bibinfo{author}{\bibfnamefont{F.}~\bibnamefont{Veloso}}, \bibnamefont{and}
  \bibinfo{author}{\bibfnamefont{A.}~\bibnamefont{Onofre}},
  \bibinfo{journal}{Eur. Phys. J.} \textbf{\bibinfo{volume}{C50}},
  \bibinfo{pages}{519} (\bibinfo{year}{2007}), \eprint{hep-ph/0605190}.

\bibitem[{\citenamefont{Englert and Spannowsky}(2015)}]{Englert:2014cva}
\bibinfo{author}{\bibfnamefont{C.}~\bibnamefont{Englert}} \bibnamefont{and}
  \bibinfo{author}{\bibfnamefont{M.}~\bibnamefont{Spannowsky}},
  \bibinfo{journal}{Phys. Lett.} \textbf{\bibinfo{volume}{B740}},
  \bibinfo{pages}{8} (\bibinfo{year}{2015}), \eprint{1408.5147}.

\bibitem[{\citenamefont{Kiyo et~al.}(2009)\citenamefont{Kiyo, Maier,
  Maierhofer, and Marquard}}]{Kiyo:2009gb}
\bibinfo{author}{\bibfnamefont{Y.}~\bibnamefont{Kiyo}},
  \bibinfo{author}{\bibfnamefont{A.}~\bibnamefont{Maier}},
  \bibinfo{author}{\bibfnamefont{P.}~\bibnamefont{Maierhofer}},
  \bibnamefont{and} \bibinfo{author}{\bibfnamefont{P.}~\bibnamefont{Marquard}},
  \bibinfo{journal}{Nucl. Phys.} \textbf{\bibinfo{volume}{B823}},
  \bibinfo{pages}{269} (\bibinfo{year}{2009}), \eprint{0907.2120}.

\bibitem[{\citenamefont{Chen et~al.}(2016)\citenamefont{Chen, Dekkers, Heisler,
  Bernreuther, and Si}}]{Chen:2016zbz}
\bibinfo{author}{\bibfnamefont{L.}~\bibnamefont{Chen}},
  \bibinfo{author}{\bibfnamefont{O.}~\bibnamefont{Dekkers}},
  \bibinfo{author}{\bibfnamefont{D.}~\bibnamefont{Heisler}},
  \bibinfo{author}{\bibfnamefont{W.}~\bibnamefont{Bernreuther}},
  \bibnamefont{and} \bibinfo{author}{\bibfnamefont{Z.-G.} \bibnamefont{Si}},
  \bibinfo{journal}{JHEP} \textbf{\bibinfo{volume}{12}}, \bibinfo{pages}{098}
  (\bibinfo{year}{2016}), \eprint{1610.07897}.

\bibitem[{\citenamefont{Klute et~al.}(2013)\citenamefont{Klute, Lafaye, Plehn,
  Rauch, and Zerwas}}]{Klute:2013cx}
\bibinfo{author}{\bibfnamefont{M.}~\bibnamefont{Klute}},
  \bibinfo{author}{\bibfnamefont{R.}~\bibnamefont{Lafaye}},
  \bibinfo{author}{\bibfnamefont{T.}~\bibnamefont{Plehn}},
  \bibinfo{author}{\bibfnamefont{M.}~\bibnamefont{Rauch}}, \bibnamefont{and}
  \bibinfo{author}{\bibfnamefont{D.}~\bibnamefont{Zerwas}},
  \bibinfo{journal}{Europhys. Lett.} \textbf{\bibinfo{volume}{101}},
  \bibinfo{pages}{51001} (\bibinfo{year}{2013}), \eprint{1301.1322}.

\bibitem[{\citenamefont{Khanpour and
  Mohammadi~Najafabadi}(2017)}]{Khanpour:2017cfq}
\bibinfo{author}{\bibfnamefont{H.}~\bibnamefont{Khanpour}} \bibnamefont{and}
  \bibinfo{author}{\bibfnamefont{M.}~\bibnamefont{Mohammadi~Najafabadi}},
  \bibinfo{journal}{Phys. Rev.} \textbf{\bibinfo{volume}{D95}},
  \bibinfo{pages}{055026} (\bibinfo{year}{2017}), \eprint{1702.00951}.

\bibitem[{\citenamefont{Vos}(2017)}]{Vos:2017ses}
\bibinfo{author}{\bibfnamefont{M.}~\bibnamefont{Vos}} (\bibinfo{year}{2017}),
  \eprint{1701.06537}.

\bibitem[{\citenamefont{Vos et~al.}(2016)}]{Vos:2016til}
\bibinfo{author}{\bibfnamefont{M.}~\bibnamefont{Vos}} \bibnamefont{et~al.}
  (\bibinfo{year}{2016}), \eprint{1604.08122}.

\bibitem[{\citenamefont{Amjad et~al.}(2015)}]{Amjad:2015mma}
\bibinfo{author}{\bibfnamefont{M.~S.} \bibnamefont{Amjad}}
  \bibnamefont{et~al.}, \bibinfo{journal}{Eur. Phys. J.}
  \textbf{\bibinfo{volume}{C75}}, \bibinfo{pages}{512} (\bibinfo{year}{2015}),
  \eprint{1505.06020}.

\bibitem[{\citenamefont{Röntsch and Schulze}(2015)}]{Rontsch:2015una}
\bibinfo{author}{\bibfnamefont{R.}~\bibnamefont{Röntsch}} \bibnamefont{and}
  \bibinfo{author}{\bibfnamefont{M.}~\bibnamefont{Schulze}},
  \bibinfo{journal}{JHEP} \textbf{\bibinfo{volume}{08}}, \bibinfo{pages}{044}
  (\bibinfo{year}{2015}), \eprint{1501.05939}.

\bibitem[{\citenamefont{Coleppa et~al.}(2017)\citenamefont{Coleppa, Kumar,
  Kumar, and Mellado}}]{Coleppa:2017rgb}
\bibinfo{author}{\bibfnamefont{B.}~\bibnamefont{Coleppa}},
  \bibinfo{author}{\bibfnamefont{M.}~\bibnamefont{Kumar}},
  \bibinfo{author}{\bibfnamefont{S.}~\bibnamefont{Kumar}}, \bibnamefont{and}
  \bibinfo{author}{\bibfnamefont{B.}~\bibnamefont{Mellado}}
  (\bibinfo{year}{2017}), \eprint{1702.03426}.

\bibitem[{\citenamefont{Cao and Yan}(2015)}]{Cao:2015qta}
\bibinfo{author}{\bibfnamefont{Q.-H.} \bibnamefont{Cao}} \bibnamefont{and}
  \bibinfo{author}{\bibfnamefont{B.}~\bibnamefont{Yan}},
  \bibinfo{journal}{Phys. Rev.} \textbf{\bibinfo{volume}{D92}},
  \bibinfo{pages}{094018} (\bibinfo{year}{2015}), \eprint{1507.06204}.

\bibitem[{\citenamefont{Grzadkowski et~al.}(2010)\citenamefont{Grzadkowski,
  Iskrzynski, Misiak, and Rosiek}}]{Grzadkowski:2010es}
\bibinfo{author}{\bibfnamefont{B.}~\bibnamefont{Grzadkowski}},
  \bibinfo{author}{\bibfnamefont{M.}~\bibnamefont{Iskrzynski}},
  \bibinfo{author}{\bibfnamefont{M.}~\bibnamefont{Misiak}}, \bibnamefont{and}
  \bibinfo{author}{\bibfnamefont{J.}~\bibnamefont{Rosiek}},
  \bibinfo{journal}{JHEP} \textbf{\bibinfo{volume}{10}}, \bibinfo{pages}{085}
  (\bibinfo{year}{2010}), \eprint{1008.4884}.

\bibitem[{\citenamefont{Cirigliano et~al.}(2016)\citenamefont{Cirigliano,
  Dekens, de~Vries, and Mereghetti}}]{Cirigliano:2016njn}
\bibinfo{author}{\bibfnamefont{V.}~\bibnamefont{Cirigliano}},
  \bibinfo{author}{\bibfnamefont{W.}~\bibnamefont{Dekens}},
  \bibinfo{author}{\bibfnamefont{J.}~\bibnamefont{de~Vries}}, \bibnamefont{and}
  \bibinfo{author}{\bibfnamefont{E.}~\bibnamefont{Mereghetti}},
  \bibinfo{journal}{Phys. Rev.} \textbf{\bibinfo{volume}{D94}},
  \bibinfo{pages}{016002} (\bibinfo{year}{2016}), \eprint{1603.03049}.

\bibitem[{\citenamefont{Alloul et~al.}(2014)\citenamefont{Alloul, Christensen,
  Degrande, Duhr, and Fuks}}]{Alloul:2013bka}
\bibinfo{author}{\bibfnamefont{A.}~\bibnamefont{Alloul}},
  \bibinfo{author}{\bibfnamefont{N.~D.} \bibnamefont{Christensen}},
  \bibinfo{author}{\bibfnamefont{C.}~\bibnamefont{Degrande}},
  \bibinfo{author}{\bibfnamefont{C.}~\bibnamefont{Duhr}}, \bibnamefont{and}
  \bibinfo{author}{\bibfnamefont{B.}~\bibnamefont{Fuks}},
  \bibinfo{journal}{Comput. Phys. Commun.} \textbf{\bibinfo{volume}{185}},
  \bibinfo{pages}{2250} (\bibinfo{year}{2014}), \eprint{1310.1921}.

\bibitem[{\citenamefont{Degrande et~al.}(2012)\citenamefont{Degrande, Duhr,
  Fuks, Grellscheid, Mattelaer, and Reiter}}]{Degrande:2011ua}
\bibinfo{author}{\bibfnamefont{C.}~\bibnamefont{Degrande}},
  \bibinfo{author}{\bibfnamefont{C.}~\bibnamefont{Duhr}},
  \bibinfo{author}{\bibfnamefont{B.}~\bibnamefont{Fuks}},
  \bibinfo{author}{\bibfnamefont{D.}~\bibnamefont{Grellscheid}},
  \bibinfo{author}{\bibfnamefont{O.}~\bibnamefont{Mattelaer}},
  \bibnamefont{and} \bibinfo{author}{\bibfnamefont{T.}~\bibnamefont{Reiter}},
  \bibinfo{journal}{Comput. Phys. Commun.} \textbf{\bibinfo{volume}{183}},
  \bibinfo{pages}{1201} (\bibinfo{year}{2012}), \eprint{1108.2040}.

\bibitem[{\citenamefont{Alwall et~al.}(2014)\citenamefont{Alwall, Frederix,
  Frixione, Hirschi, Maltoni, Mattelaer, Shao, Stelzer, Torrielli, and
  Zaro}}]{Alwall:2014hca}
\bibinfo{author}{\bibfnamefont{J.}~\bibnamefont{Alwall}},
  \bibinfo{author}{\bibfnamefont{R.}~\bibnamefont{Frederix}},
  \bibinfo{author}{\bibfnamefont{S.}~\bibnamefont{Frixione}},
  \bibinfo{author}{\bibfnamefont{V.}~\bibnamefont{Hirschi}},
  \bibinfo{author}{\bibfnamefont{F.}~\bibnamefont{Maltoni}},
  \bibinfo{author}{\bibfnamefont{O.}~\bibnamefont{Mattelaer}},
  \bibinfo{author}{\bibfnamefont{H.~S.} \bibnamefont{Shao}},
  \bibinfo{author}{\bibfnamefont{T.}~\bibnamefont{Stelzer}},
  \bibinfo{author}{\bibfnamefont{P.}~\bibnamefont{Torrielli}},
  \bibnamefont{and} \bibinfo{author}{\bibfnamefont{M.}~\bibnamefont{Zaro}},
  \bibinfo{journal}{JHEP} \textbf{\bibinfo{volume}{07}}, \bibinfo{pages}{079}
  (\bibinfo{year}{2014}), \eprint{1405.0301}.

\bibitem[{\citenamefont{Hahn}(2001)}]{Hahn:2000kx}
\bibinfo{author}{\bibfnamefont{T.}~\bibnamefont{Hahn}},
  \bibinfo{journal}{Comput. Phys. Commun.} \textbf{\bibinfo{volume}{140}},
  \bibinfo{pages}{418} (\bibinfo{year}{2001}), \eprint{hep-ph/0012260}.

\bibitem[{\citenamefont{Hahn and Perez-Victoria}(1999)}]{Hahn:1998yk}
\bibinfo{author}{\bibfnamefont{T.}~\bibnamefont{Hahn}} \bibnamefont{and}
  \bibinfo{author}{\bibfnamefont{M.}~\bibnamefont{Perez-Victoria}},
  \bibinfo{journal}{Comput. Phys. Commun.} \textbf{\bibinfo{volume}{118}},
  \bibinfo{pages}{153} (\bibinfo{year}{1999}), \eprint{hep-ph/9807565}.

\bibitem[{\citenamefont{James and Roos}(1975)}]{James:1975dr}
\bibinfo{author}{\bibfnamefont{F.}~\bibnamefont{James}} \bibnamefont{and}
  \bibinfo{author}{\bibfnamefont{M.}~\bibnamefont{Roos}},
  \bibinfo{journal}{Comput. Phys. Commun.} \textbf{\bibinfo{volume}{10}},
  \bibinfo{pages}{343} (\bibinfo{year}{1975}).

\bibitem[{\citenamefont{Buckley et~al.}(2010)\citenamefont{Buckley, Hoeth,
  Lacker, Schulz, and von Seggern}}]{Buckley:2009bj}
\bibinfo{author}{\bibfnamefont{A.}~\bibnamefont{Buckley}},
  \bibinfo{author}{\bibfnamefont{H.}~\bibnamefont{Hoeth}},
  \bibinfo{author}{\bibfnamefont{H.}~\bibnamefont{Lacker}},
  \bibinfo{author}{\bibfnamefont{H.}~\bibnamefont{Schulz}}, \bibnamefont{and}
  \bibinfo{author}{\bibfnamefont{J.~E.} \bibnamefont{von Seggern}},
  \bibinfo{journal}{Eur. Phys. J.} \textbf{\bibinfo{volume}{C65}},
  \bibinfo{pages}{331} (\bibinfo{year}{2010}), \eprint{0907.2973}.

\bibitem[{\citenamefont{Maltoni et~al.}(2016)\citenamefont{Maltoni, Pagani, and
  Tsinikos}}]{Maltoni:2015ena}
\bibinfo{author}{\bibfnamefont{F.}~\bibnamefont{Maltoni}},
  \bibinfo{author}{\bibfnamefont{D.}~\bibnamefont{Pagani}}, \bibnamefont{and}
  \bibinfo{author}{\bibfnamefont{I.}~\bibnamefont{Tsinikos}},
  \bibinfo{journal}{JHEP} \textbf{\bibinfo{volume}{02}}, \bibinfo{pages}{113}
  (\bibinfo{year}{2016}), \eprint{1507.05640}.

\bibitem[{\citenamefont{Aaboud et~al.}(2017)}]{Aaboud:2016xve}
\bibinfo{author}{\bibfnamefont{M.}~\bibnamefont{Aaboud}} \bibnamefont{et~al.}
  (\bibinfo{collaboration}{ATLAS}), \bibinfo{journal}{Eur. Phys. J.}
  \textbf{\bibinfo{volume}{C77}}, \bibinfo{pages}{40} (\bibinfo{year}{2017}),
  \eprint{1609.01599}.

\bibitem[{CMS(2016)}]{CMS-PAS-TOP-16-017}
\bibinfo{type}{Tech. Rep.} \bibinfo{number}{CMS-PAS-TOP-16-017},
  \bibinfo{institution}{CERN}, \bibinfo{address}{Geneva}
  (\bibinfo{year}{2016}), \urlprefix\url{https://cds.cern.ch/record/2205283}.

\bibitem[{\citenamefont{Englert et~al.}(2016)\citenamefont{Englert, Moore,
  Nordström, and Russell}}]{Englert:2016aei}
\bibinfo{author}{\bibfnamefont{C.}~\bibnamefont{Englert}},
  \bibinfo{author}{\bibfnamefont{L.}~\bibnamefont{Moore}},
  \bibinfo{author}{\bibfnamefont{K.}~\bibnamefont{Nordström}},
  \bibnamefont{and} \bibinfo{author}{\bibfnamefont{M.}~\bibnamefont{Russell}},
  \bibinfo{journal}{Phys. Lett.} \textbf{\bibinfo{volume}{B763}},
  \bibinfo{pages}{9} (\bibinfo{year}{2016}), \eprint{1607.04304}.

\bibitem[{\citenamefont{Boogert and Miller}(2002)}]{Boogert:2002jr}
\bibinfo{author}{\bibfnamefont{S.~T.} \bibnamefont{Boogert}} \bibnamefont{and}
  \bibinfo{author}{\bibfnamefont{D.~J.} \bibnamefont{Miller}}, in
  \emph{\bibinfo{booktitle}{{Linear colliders. Proceedings, International
  Workshop on physics and experiments with future electron-positron linear
  colliders, LCWS 2002, Seogwipo, Jeju Island, Korea, August 26-30, 2002}}}
  (\bibinfo{year}{2002}), pp. \bibinfo{pages}{509--516},
  \eprint{hep-ex/0211021},
  \urlprefix\url{http://alice.cern.ch/format/showfull?sysnb=2349913}.

\bibitem[{\citenamefont{Chokoufé~Nejad
  et~al.}(2016)\citenamefont{Chokoufé~Nejad, Kilian, Lindert, Pozzorini,
  Reuter, and Weiss}}]{Nejad:2016bci}
\bibinfo{author}{\bibfnamefont{B.}~\bibnamefont{Chokoufé~Nejad}},
  \bibinfo{author}{\bibfnamefont{W.}~\bibnamefont{Kilian}},
  \bibinfo{author}{\bibfnamefont{J.~M.} \bibnamefont{Lindert}},
  \bibinfo{author}{\bibfnamefont{S.}~\bibnamefont{Pozzorini}},
  \bibinfo{author}{\bibfnamefont{J.}~\bibnamefont{Reuter}}, \bibnamefont{and}
  \bibinfo{author}{\bibfnamefont{C.}~\bibnamefont{Weiss}},
  \bibinfo{journal}{JHEP} \textbf{\bibinfo{volume}{12}}, \bibinfo{pages}{075}
  (\bibinfo{year}{2016}), \eprint{1609.03390}.

\bibitem[{\citenamefont{Fleischer et~al.}(2003)\citenamefont{Fleischer, Leike,
  Riemann, and Werthenbach}}]{Fleischer:2003kk}
\bibinfo{author}{\bibfnamefont{J.}~\bibnamefont{Fleischer}},
  \bibinfo{author}{\bibfnamefont{A.}~\bibnamefont{Leike}},
  \bibinfo{author}{\bibfnamefont{T.}~\bibnamefont{Riemann}}, \bibnamefont{and}
  \bibinfo{author}{\bibfnamefont{A.}~\bibnamefont{Werthenbach}},
  \bibinfo{journal}{Eur. Phys. J.} \textbf{\bibinfo{volume}{C31}},
  \bibinfo{pages}{37} (\bibinfo{year}{2003}), \eprint{hep-ph/0302259}.

\bibitem[{\citenamefont{Gao and Zhu}(2014{\natexlab{a}})}]{Gao:2014nva}
\bibinfo{author}{\bibfnamefont{J.}~\bibnamefont{Gao}} \bibnamefont{and}
  \bibinfo{author}{\bibfnamefont{H.~X.} \bibnamefont{Zhu}},
  \bibinfo{journal}{Phys. Rev.} \textbf{\bibinfo{volume}{D90}},
  \bibinfo{pages}{114022} (\bibinfo{year}{2014}{\natexlab{a}}),
  \eprint{1408.5150}.

\bibitem[{\citenamefont{Bernreuther et~al.}(2006)\citenamefont{Bernreuther,
  Bonciani, Gehrmann, Heinesch, Leineweber, Mastrolia, and
  Remiddi}}]{Bernreuther:2006vp}
\bibinfo{author}{\bibfnamefont{W.}~\bibnamefont{Bernreuther}},
  \bibinfo{author}{\bibfnamefont{R.}~\bibnamefont{Bonciani}},
  \bibinfo{author}{\bibfnamefont{T.}~\bibnamefont{Gehrmann}},
  \bibinfo{author}{\bibfnamefont{R.}~\bibnamefont{Heinesch}},
  \bibinfo{author}{\bibfnamefont{T.}~\bibnamefont{Leineweber}},
  \bibinfo{author}{\bibfnamefont{P.}~\bibnamefont{Mastrolia}},
  \bibnamefont{and} \bibinfo{author}{\bibfnamefont{E.}~\bibnamefont{Remiddi}},
  \bibinfo{journal}{Nucl. Phys.} \textbf{\bibinfo{volume}{B750}},
  \bibinfo{pages}{83} (\bibinfo{year}{2006}), \eprint{hep-ph/0604031}.

\bibitem[{\citenamefont{Gao and Zhu}(2014{\natexlab{b}})}]{Gao:2014eea}
\bibinfo{author}{\bibfnamefont{J.}~\bibnamefont{Gao}} \bibnamefont{and}
  \bibinfo{author}{\bibfnamefont{H.~X.} \bibnamefont{Zhu}},
  \bibinfo{journal}{Phys. Rev. Lett.} \textbf{\bibinfo{volume}{113}},
  \bibinfo{pages}{262001} (\bibinfo{year}{2014}{\natexlab{b}}),
  \eprint{1410.3165}.

\bibitem[{\citenamefont{Hikasa}(1986)}]{Hikasa:1985qi}
\bibinfo{author}{\bibfnamefont{K.-i.} \bibnamefont{Hikasa}},
  \bibinfo{journal}{Phys. Rev.} \textbf{\bibinfo{volume}{D33}},
  \bibinfo{pages}{3203} (\bibinfo{year}{1986}).

\bibitem[{\citenamefont{Moortgat-Pick et~al.}(2008)}]{MoortgatPick:2005cw}
\bibinfo{author}{\bibfnamefont{G.}~\bibnamefont{Moortgat-Pick}}
  \bibnamefont{et~al.}, \bibinfo{journal}{Phys. Rept.}
  \textbf{\bibinfo{volume}{460}}, \bibinfo{pages}{131} (\bibinfo{year}{2008}),
  \eprint{hep-ph/0507011}.

\bibitem[{\citenamefont{Amjad et~al.}(2013)\citenamefont{Amjad, Boronat,
  Frisson, Garcia, Poschl, Ros, Richard, Rouene, Femenia, and
  Vos}}]{Amjad:2013tlv}
\bibinfo{author}{\bibfnamefont{M.~S.} \bibnamefont{Amjad}},
  \bibinfo{author}{\bibfnamefont{M.}~\bibnamefont{Boronat}},
  \bibinfo{author}{\bibfnamefont{T.}~\bibnamefont{Frisson}},
  \bibinfo{author}{\bibfnamefont{I.}~\bibnamefont{Garcia}},
  \bibinfo{author}{\bibfnamefont{R.}~\bibnamefont{Poschl}},
  \bibinfo{author}{\bibfnamefont{E.}~\bibnamefont{Ros}},
  \bibinfo{author}{\bibfnamefont{F.}~\bibnamefont{Richard}},
  \bibinfo{author}{\bibfnamefont{J.}~\bibnamefont{Rouene}},
  \bibinfo{author}{\bibfnamefont{P.~R.} \bibnamefont{Femenia}},
  \bibnamefont{and} \bibinfo{author}{\bibfnamefont{M.}~\bibnamefont{Vos}}
  (\bibinfo{year}{2013}), \eprint{1307.8102}.

\bibitem[{\citenamefont{Boland et~al.}(2016)}]{CLIC:2016zwp}
\bibinfo{author}{\bibfnamefont{M.~J.} \bibnamefont{Boland}}
  \bibnamefont{et~al.} (\bibinfo{collaboration}{CLICdp, CLIC})
  (\bibinfo{year}{2016}), \eprint{1608.07537}.

\bibitem[{\citenamefont{Abramowicz et~al.}(2016)}]{Abramowicz:2016zbo}
\bibinfo{author}{\bibfnamefont{H.}~\bibnamefont{Abramowicz}}
  \bibnamefont{et~al.} (\bibinfo{year}{2016}), \eprint{1608.07538}.

\bibitem[{\citenamefont{Ellis et~al.}(2017)\citenamefont{Ellis, Roloff, Sanz,
  and You}}]{Ellis:2017kfi}
\bibinfo{author}{\bibfnamefont{J.}~\bibnamefont{Ellis}},
  \bibinfo{author}{\bibfnamefont{P.}~\bibnamefont{Roloff}},
  \bibinfo{author}{\bibfnamefont{V.}~\bibnamefont{Sanz}}, \bibnamefont{and}
  \bibinfo{author}{\bibfnamefont{T.}~\bibnamefont{You}} (\bibinfo{year}{2017}),
  \eprint{1701.04804}.

\bibitem[{\citenamefont{Berthier and Trott}(2015)}]{Berthier:2015oma}
\bibinfo{author}{\bibfnamefont{L.}~\bibnamefont{Berthier}} \bibnamefont{and}
  \bibinfo{author}{\bibfnamefont{M.}~\bibnamefont{Trott}},
  \bibinfo{journal}{JHEP} \textbf{\bibinfo{volume}{05}}, \bibinfo{pages}{024}
  (\bibinfo{year}{2015}), \eprint{1502.02570}.

\bibitem[{\citenamefont{Ghezzi et~al.}(2015)\citenamefont{Ghezzi,
  Gomez-Ambrosio, Passarino, and Uccirati}}]{Ghezzi:2015vva}
\bibinfo{author}{\bibfnamefont{M.}~\bibnamefont{Ghezzi}},
  \bibinfo{author}{\bibfnamefont{R.}~\bibnamefont{Gomez-Ambrosio}},
  \bibinfo{author}{\bibfnamefont{G.}~\bibnamefont{Passarino}},
  \bibnamefont{and} \bibinfo{author}{\bibfnamefont{S.}~\bibnamefont{Uccirati}},
  \bibinfo{journal}{JHEP} \textbf{\bibinfo{volume}{07}}, \bibinfo{pages}{175}
  (\bibinfo{year}{2015}), \eprint{1505.03706}.

\bibitem[{\citenamefont{Baak et~al.}(2014)\citenamefont{Baak, Cuth, Haller,
  Hoecker, Kogler, Mönig, Schott, and Stelzer}}]{Baak:2014ora}
\bibinfo{author}{\bibfnamefont{M.}~\bibnamefont{Baak}},
  \bibinfo{author}{\bibfnamefont{J.}~\bibnamefont{Cuth}},
  \bibinfo{author}{\bibfnamefont{J.}~\bibnamefont{Haller}},
  \bibinfo{author}{\bibfnamefont{A.}~\bibnamefont{Hoecker}},
  \bibinfo{author}{\bibfnamefont{R.}~\bibnamefont{Kogler}},
  \bibinfo{author}{\bibfnamefont{K.}~\bibnamefont{Mönig}},
  \bibinfo{author}{\bibfnamefont{M.}~\bibnamefont{Schott}}, \bibnamefont{and}
  \bibinfo{author}{\bibfnamefont{J.}~\bibnamefont{Stelzer}}
  (\bibinfo{collaboration}{Gfitter Group}), \bibinfo{journal}{Eur. Phys. J.}
  \textbf{\bibinfo{volume}{C74}}, \bibinfo{pages}{3046} (\bibinfo{year}{2014}),
  \eprint{1407.3792}.

\bibitem[{\citenamefont{Peskin and Takeuchi}(1990)}]{Peskin:1990zt}
\bibinfo{author}{\bibfnamefont{M.~E.} \bibnamefont{Peskin}} \bibnamefont{and}
  \bibinfo{author}{\bibfnamefont{T.}~\bibnamefont{Takeuchi}},
  \bibinfo{journal}{Phys. Rev. Lett.} \textbf{\bibinfo{volume}{65}},
  \bibinfo{pages}{964} (\bibinfo{year}{1990}).

\bibitem[{\citenamefont{Peskin and Takeuchi}(1992)}]{Peskin:1991sw}
\bibinfo{author}{\bibfnamefont{M.~E.} \bibnamefont{Peskin}} \bibnamefont{and}
  \bibinfo{author}{\bibfnamefont{T.}~\bibnamefont{Takeuchi}},
  \bibinfo{journal}{Phys. Rev.} \textbf{\bibinfo{volume}{D46}},
  \bibinfo{pages}{381} (\bibinfo{year}{1992}).

\bibitem[{\citenamefont{Ross and Veltman}(1975)}]{Ross:1975fq}
\bibinfo{author}{\bibfnamefont{D.~A.} \bibnamefont{Ross}} \bibnamefont{and}
  \bibinfo{author}{\bibfnamefont{M.~J.~G.} \bibnamefont{Veltman}},
  \bibinfo{journal}{Nucl. Phys.} \textbf{\bibinfo{volume}{B95}},
  \bibinfo{pages}{135} (\bibinfo{year}{1975}).

\bibitem[{\citenamefont{Grojean et~al.}(2013)\citenamefont{Grojean, Jenkins,
  Manohar, and Trott}}]{Grojean:2013kd}
\bibinfo{author}{\bibfnamefont{C.}~\bibnamefont{Grojean}},
  \bibinfo{author}{\bibfnamefont{E.~E.} \bibnamefont{Jenkins}},
  \bibinfo{author}{\bibfnamefont{A.~V.} \bibnamefont{Manohar}},
  \bibnamefont{and} \bibinfo{author}{\bibfnamefont{M.}~\bibnamefont{Trott}},
  \bibinfo{journal}{JHEP} \textbf{\bibinfo{volume}{04}}, \bibinfo{pages}{016}
  (\bibinfo{year}{2013}), \eprint{1301.2588}.

\bibitem[{\citenamefont{Barbieri et~al.}(2004)\citenamefont{Barbieri, Pomarol,
  Rattazzi, and Strumia}}]{Barbieri:2004qk}
\bibinfo{author}{\bibfnamefont{R.}~\bibnamefont{Barbieri}},
  \bibinfo{author}{\bibfnamefont{A.}~\bibnamefont{Pomarol}},
  \bibinfo{author}{\bibfnamefont{R.}~\bibnamefont{Rattazzi}}, \bibnamefont{and}
  \bibinfo{author}{\bibfnamefont{A.}~\bibnamefont{Strumia}},
  \bibinfo{journal}{Nucl. Phys.} \textbf{\bibinfo{volume}{B703}},
  \bibinfo{pages}{127} (\bibinfo{year}{2004}), \eprint{hep-ph/0405040}.

\bibitem[{\citenamefont{Abdallah et~al.}(2004)}]{Abdallah:2003gp}
\bibinfo{author}{\bibfnamefont{J.}~\bibnamefont{Abdallah}} \bibnamefont{et~al.}
  (\bibinfo{collaboration}{DELPHI}), \bibinfo{journal}{Eur. Phys. J.}
  \textbf{\bibinfo{volume}{C34}}, \bibinfo{pages}{109} (\bibinfo{year}{2004}),
  \eprint{hep-ex/0403041}.

\end{thebibliography}
\end{document}